\documentclass[final,3p,11pt,times]{elsarticle}
\usepackage[T1]{fontenc}
\usepackage{float}
\usepackage[utf8]{inputenc}
\usepackage{algorithm}
\usepackage{algpseudocode}
\usepackage{helvet}
\usepackage{amsmath}
\usepackage{subcaption}
\usepackage{amssymb}
\usepackage{mathrsfs}
\usepackage{natbib}
\usepackage{color}
\usepackage{fancyhdr} 
\usepackage{colortbl} 
\usepackage{graphicx}
\usepackage{array} 
\usepackage{framed}
\usepackage{url}

\usepackage{pgfplots}
\pgfplotsset{compat=1.18}
\usepackage{tcolorbox}
\usepackage{multirow}

\DeclareMathOperator*{\argmax}{arg\,max}

\begin{document}
\begin{frontmatter}

\title{
Analysis of trade-offs in urban heat mitigation using a Bayesian Optimization framework for an urban canopy layer model
}
\author[tubs-ddm]{R. Walter\corref{cor1}}
\ead{r.walter@tu-braunschweig.de}

\author[tubs-geo]{J. Gelhaus}
\author[tubs-ddm]{D. Anton}
\author[tubs-ddm]{H. Wessels}
\author[tubs-geo]{S. Weber}

\cortext[cor1]{Corresponding author}

\address[tubs-ddm]{Division Data-driven Modeling of Mechanical Systems, Institute of Applied Mechanics, Technische Universit\"at Braunschweig, Pockelsstra\ss e 3, 38106 Braunschweig, Germany}
\address[tubs-geo]{Climatology and Environmental Meteorology, Institute of Geoecology, Technische Universit\"at Braunschweig, Langer Kamp 19c, 38106 Braunschweig, Germany}

\maketitle
\thispagestyle{empty}

\begin{abstract}
To mitigate the challenges of climate change and intensifying heat stress in urban areas, local adaptation strategies are discussed and introduced in cities worldwide. To understand processes and potential trade-offs of these strategies a Bayesian optimization and surrogate modeling framework was employed to investigate urban parameter ranges of heat mitigation strategies with focus on three thermal metrics: daytime air temperature, Universal Thermal Climate Index (UTCI), and nighttime air temperature. Based on an urban street canyon configuration, it was shown that heat mitigation measures that reduce daytime air temperature and UTCI are often associated with higher nighttime temperatures. This results in a curved Pareto front that reflects the trade-off between daytime and nighttime thermal comfort.  Under identical forcing conditions, different urban configurations, varying in geometry, vegetation, and surface characteristics, are shown to alter peak canyon air temperature by up to 5.2~$^\circ$C during the day and 2.6~$^\circ$C at night, while UTCI varies by up to 7.9~$^\circ$C, demonstrating that favorable urban configurations can substantially mitigate microclimatic heat stress. 
These findings suggest that combining multi-objective Bayesian optimization and surrogate modeling can help bridge the gap between computationally intensive climate simulations and practical decision-making in urban planning. An interactive visualization tool was developed to explore these trade-offs, making the often opposing relationships between urban  parameters and the three thermal metrics directly accessible to planners.
\end{abstract}


\end{frontmatter}

\tableofcontents
\newpage
\section{Introduction}
The increasing frequency of extreme weather events, such as heat waves, driven by progressing climate change, poses significant threats to the welfare of urban inhabitants \citep{lee_ipcc_2023, luthi_rapid_2023}. Because of their dense and higher building structures and heat-retaining materials, urban areas exhibit altered surface radiation and energy balances compared to rural areas \citep{kuttler2023}. Heat is stored within the building structures during the day and released at night, extenuating the night-time cooling of cities. Coupled with strong surface heating during the day, urban residents are often exposed to significant heat stress during the summer months which results in an increased heat-related morbidity and mortality \citep{masselot_estimating_2025}.
For instance, the German federal public health institute (Robert Koch-Institut) estimated that the June 2026 heat wave resulted in about 5,100 heat-related deaths in Germany \citep{robert_koch-institut_wochenbericht_2026}. About 67 \% of the world’s population is predicted to reside in cities by 2050 and urban areas are expected to expand and densify further \citep{UN2025Urbanization}. Therefore, targeted adaptation measures to urban overheating are crucial to protect a major share of the world's population.

A range of heat mitigation measures has been investigated, including highly reflective surfaces such as 'super cool roofs', shading structures, and green infrastructure such as street trees and vegetated areas \citep{kumar_urban_2024, FAYMONVILLE2026102956, elnabawi_super_2023, markolf-2026_NEE_greenRoof}. Highly reflective surfaces reduce the amount of solar radiation absorbed by urban structures, while green infrastructure counteracts urban heating through shade and evapotranspiration \citep{kumar_urban_2024}. However, the evaluation of these measures through modeling and observations has revealed that their effectiveness is highly sensitive to local context, including urban morphology, land use, and the prevailing mesoscale climate, motivating the need for flexible and locally applicable modeling frameworks.

Urban canopy models resolving the energy balance fill this role by simulating the microclimate within a simplified urban geometry, balancing physical realism with computational tractability \citep{masson_physically-based_2000}. Compared to full computational fluid dynamics approaches, they sacrifice some small-scale, building resolving resolution \citep{mateen_large_2025}, but remain well-suited for evaluating localized adaptation strategies under realistic meteorological forcing. The Urban Tethys-Chloris model (UT\&C) represents a particularly capable example, explicitly resolving ground vegetation, street trees, and their effects on transpirative and aerodynamic exchange \citep{meili2020utc, meili2021trees}. It can further be coupled to a building energy model to capture HVAC feedback on the urban microclimate \citep{meili_modeling_2025}.

Numerous studies have evaluated urban heat mitigation strategies using urban canopy models \citep{krayenhoff_cooling_2021}. However, by investigating specific adaptation scenarios for a certain study area, most studies only compare the impact of a small, specific range of adaptation measures to the status quo. Covering entire parameter ranges of heat mitigation measures, such as tree size and building surface parameters, provides a systematic assessment of how their effectiveness varies across their full range of implementation. Moreover, by focusing on the optimization of a single thermal metric such as air temperature, potential trade-offs between competing objectives are not fully addressed. For instance, increasing surface albedo can reduce absorbed solar radiation but may simultaneously raise exposure to reflected shortwave radiation for pedestrians \citep{schrijvers_effect_2016}. Similarly, street trees provide daytime shade and evapotranspirative cooling but may inhibit canyon ventilation and reduce longwave radiative cooling at night \citep{meili2021trees}. These interacting effects highlight the complexity of urban heat mitigation and the need for approaches capable of systematically exploring large parameter spaces across multiple thermal objectives.

This study addresses this research gap by systematically investigating daytime thermal comfort and nighttime cooling within urban street canyons during heat wave conditions using UT\&C. Using observational data from two street canyons in Braunschweig, Germany, we verify the ability of the model to reproduce the effect of contrasting geometry and vegetation on urban climate. We then apply a multi-objective Bayesian optimization (BO) framework to identify urban configurations representing favorable trade-offs between the considered heat stress indicators. Rather than evaluating distinct heat mitigation scenarios, the framework enables exploration of entire parameter spaces of urban design variables. The impact of the design variables is additionally visualized through an interactive slider tool, making the effect of individual parameters more accessible for planning applications to enable a locally customized implementation of heat mitigation strategies.

\section{Methodology}

Several quantities are of interest when assessing urban heat stress \citep{lo_optimal_2023}. Air temperature is the most widely used metric, due to its straightforward measurability and reproducibility. The urban–rural difference in nighttime air temperature is also commonly studied as an indicator of urban heat island intensity \citep{OKE1973769}. Human-biometeorological indices such as the Universal Thermal Climate Index (UTCI) \citep{brodeDerivingOperationalProcedure2012, jendritzky_utciwhy_2012} extend beyond air temperature by integrating humidity, mean radiant temperature, and wind speed, and have gained increasing relevance in recent years \citep{Krueger2021, lo_optimal_2023, shen_parametric_2026}. However, most studies on urban climate heat mitigation or heat exposure focus on a single thermal metric, potentially overlooking trade-offs between air temperature reduction and actual thermal comfort. To capture such dynamics, we jointly investigate three metrics: peak daytime air temperature to reflect peak heat exposure during daytime, minimum nighttime air temperature to characterize nocturnal cooling, both simulated at 2 m above ground, and UTCI as a heat stress assessment indicator. This selection reflects the potentially distinct effects of heat mitigation measures across daytime and nighttime conditions, as well as between air temperature and perceived thermal comfort.

The three metrics are defined over a simulation interval $t\in\left[0,\, t_{\mathrm{obs}}\right]$, where $t_{\mathrm{obs}}$ denotes the final simulation time.
\begin{itemize}
    \item The daytime peak temperature $T_\mathrm{peak}$ is the global maximum of the air temperature time series $T_\mathrm{2\text{m}}(t;\boldsymbol{\beta})$, where $\boldsymbol{\beta}$ denotes the vector of parameters characterizing street canyon morphology:
    \begin{equation}\label{eq:peak}
        T_\mathrm{peak} (\boldsymbol{\beta}) = \max \, T_{2\text{m}}(t;\boldsymbol{\beta}), \quad t\in\left[0, t_\mathrm{obs}\right]
    \end{equation}
    \item The nighttime minimum temperature $T_\mathrm{\text{night}}$ is the minimum nocturnal temperature following the hottest day in the simulation period that is defined as the day on which the atmospheric temperature $T_\mathrm{\text{atm}}$ reaches its maximum:
    \begin{equation}\label{eq:Night}
    \begin{aligned}
        &t_\mathrm{\text{hot}} = \argmax_{t} \, T_\mathrm{atm}(t), \quad t\in\left[0, t_\mathrm{obs}\right] \\
        &T_\mathrm{night}(\boldsymbol{\beta}) =
        \min T_{2\text{m}}(t;\boldsymbol{\beta}), \quad t\in\left[ t_\mathrm{\text{hot}}, \,  t_\mathrm{\text{hot}} +16\text{h}\right]
    \end{aligned}
    \end{equation}
    \item The $\text{UTCI}_\mathrm{\text{peak}}$ is the global maximum in $\text{UTCI}_\mathrm{1.1\text{m}}$ during the observation time $t\in\left[0,\, t_\mathrm{\text{obs}}\right]$. It takes temperature $T_\mathrm{2\text{m}}$, relative humidity $RH_\mathrm{\text{2m}}$, mean radiant temperature $T_R$ and wind speed at person height $v_\mathrm{1.1\text{m}}$ into account and is simulated at a height of 1.1 m above ground level:
    \begin{equation}\label{eq:UTCIpeak}
        \text{UTCI}_\mathrm{peak}(\boldsymbol{\beta}) = \max \,  \text{UTCI}_\mathrm{1.1\text{m}}(t, T_\mathrm{2\text{m}}, RH_{\mathrm{2\text{m}}},T_\mathrm{R} ,v_\mathrm{1.1\text{m}}; \boldsymbol{\beta}) , \quad t\in\left[0, t_\mathrm{obs}\right]
    \end{equation}
\end{itemize}
For as-built conditions, the metrics can be obtained from meteorological observations: $T_\mathrm{peak}$ and $T_\mathrm{night}$ follow directly from air temperature measurements, while $\text{UTCI}_\mathrm{peak}$ additionally requires relative humidity (RH), wind speed, and radiation metrics. 

These metrics are computed using the UT\&C model (Section~\ref{sec:UTC}) and embedded within a multi-objective BO framework that jointly accounts for all three objectives (Section~\ref{sec:pareto}).

\subsection{Study Area and Data Acquisition}
\label{sec:StudyArea}

\begin{figure}[htbp]
    \centering
    \includegraphics[height=0.4\textheight]{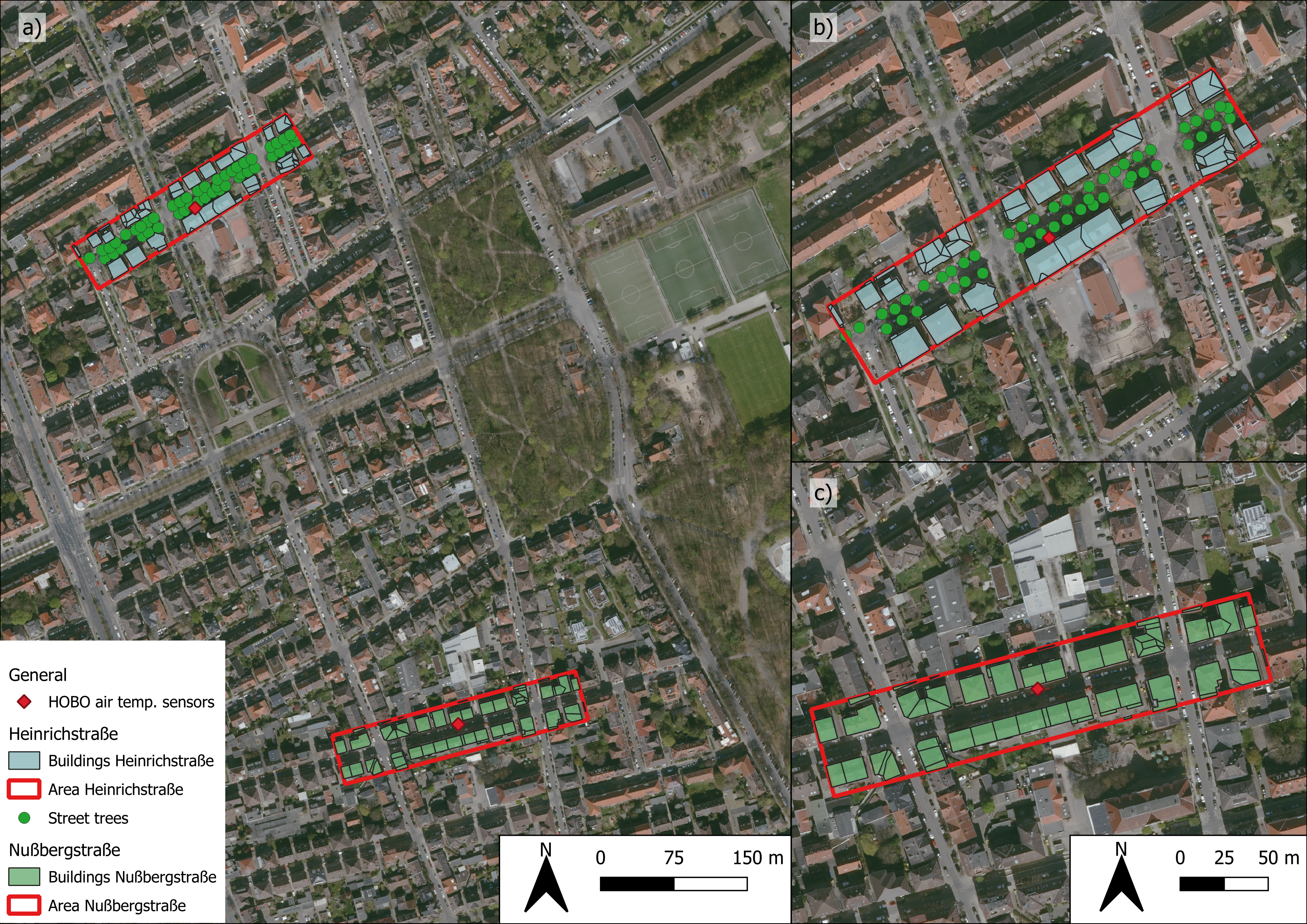}
    \caption{a) Map of the urban area of Heinrichstraße (HST) and Nußbergstraße (NST). b) HST area of interest, including the LoD2 building shapes, street trees, and placement of the air temperature sensor. c) NST area of interest, including the LoD2 building shapes, street trees, and placement of the air temperature sensor.}
    \label{fig:HST_Street_area}
\end{figure}

Two urban street canyons in Braunschweig, Germany were analyzed in this study (Figure~\ref{fig:HST_Street_area}). Heinrichstraße (HST) and Nußbergstraße (NST) are in close proximity and share a similar west–east orientation, with HST being oriented slightly farther north. The streets are located in relatively close proximity to an urban park (Prinz-Albrecht Park) towards the east with an extension of 46.5 ha \citep{braunschweig_prinzpark}. Measurements of air temperature and humidity were conducted in both street canyons from July 1 to October 31, 2021. Temperature and humidity sensors (HOBO U23 Pro v2 External Temperature/Relative Humidity Data Logger – U23–002, Onset Computer Corporation) were installed on a tree in HST and on a streetlight pole in NST, both at a height of 3 m above ground level to protect the sensors from vandalism. The sensors measured at a 5 minute temporal resolution. Atmospheric forcing temperature $T_\mathrm{\text{atm}}$ is collected from the German Weather Service station located in a rural area north-east of the central urban area of Braunschweig with a distance of about 6.85 and 7.30 km to HST and NST respectively (Station 00662; \citealp{dwd_cdc_climate_data}). Larger gaps in the wind speed forcing data that occurred within the study period were filled with wind speed data measured at the Institute of Geoecology, Technische Universität Braunschweig which is located at about 0.70 and 1.30 km from HST and NST, respectively.

Geometric properties such as building height and width of the street canyon were derived from level of detail 2 (LoD2) building geodata, provided by the State Office for Geoinformation and Surveying of Lower Saxony \citep{lgln2024_lod2}. The calculation of the average building height $H_\mathrm{c}$ was conducted for a rectangle area spanning about 257 m x 51 m and was weighted by the individual roof area of the buildings. The average roof width $W_\mathrm{r}$ was determined perpendicular to the street orientation. The derived geometric properties indicate slightly higher and larger buildings in HST, while NST features a more narrow canyon (Table~\ref{tab:utc_eval_params}). HST is characterized by a higher share of vegetation with a large number of street trees whereas NST does not entail any street trees or roadside vegetation. Both streets can be assigned to the local climate zone 2 'Compact midrise' \citep{stewart_local_2012}. Average tree radius $R_\mathrm{tree}$ was determined from LiDAR-based tree data supplied by the IOER Research Data Centre \citep{muenzinger2025semantic}. Only trees that were located in the canyon between the buildings were considered. Trees located in gardens behind buildings or in perpendicular roads were not taken into consideration to imitate the model configuration as accurately as possible. The fraction of ground vegetation $f_{\mathrm{veg,G}}$ was visually estimated from orthophotos and in-situ observations. 

As there was no detailed data available on surface albedo for walls and roofs ($\alpha_\mathrm{w}$ and $\alpha_\mathrm{r}$) as well as thermal conductivity and volumetric heat capacity of walls ($\lambda_\mathrm{w}$ and $c_\mathrm{v,s,W}$) and impervious ground surfaces ($\lambda_\mathrm{dry,G}$ and $c_\mathrm{v,s,G}$), the parameters were adopted from the Zurich parameterization presented by \citet{meili2020utc}.

\begin{table}[htbp]
\centering
\caption{Geometric and vegetation parameters of the individual streets used for model evaluation}
\label{tab:utc_eval_params}
\begin{tabular}{lccc}
\hline
\textbf{Parameter} & \textbf{Symbol} & \textbf{HST} & \textbf{NST} \\
\hline
{Location} &      & 52.270905, 10.542017 & 52.266003, 10.545799 \\
{Canyon orientation [$^\circ$]} &      & 59.85 & 76.18 \\
{Height canyon [m]} & $H_\mathrm{\text{c}}$     & 20.56 & 18.36 \\
{Width canyon [m]}  & $W_\mathrm{\text{c}}$    & 22.50 &  13.61 \\
{Width roof [m]} & $W_\mathrm{\text{r}}$ & 14.43 &  13.70 \\
{Radius tree [m]}  & $R_\mathrm{\text{tree}}$ & 4.15 &  - \\
{Height tree [m]}    & $H_\mathrm{\text{tree}}$ &  12.65  &  - \\
{Distance tree to wall [m]}  &$D_\mathrm{\text{tree}}$ & 0.5 + $R_\mathrm{\text{tree}}$  &  - \\
{Fraction vegetated ground [-]}    &$f_\mathrm{\text{veg,G}}$    & 0.2  & 0.0  \\
{Fraction bare ground [-]}    &$f_\mathrm{\text{bare,G}}$    & 0.1  & 0.1  \\
{Fraction impervious ground [-]} &$f_\mathrm{\text{imp,G}}$    & 0.7  & 0.9  \\
\hline
\end{tabular}
\end{table}

\subsection{Urban Tethys–Chloris Model (UT\&C)}
\label{sec:UTC}

\begin{figure}[htbp]
    \centering
    \includegraphics[width=0.8\linewidth]{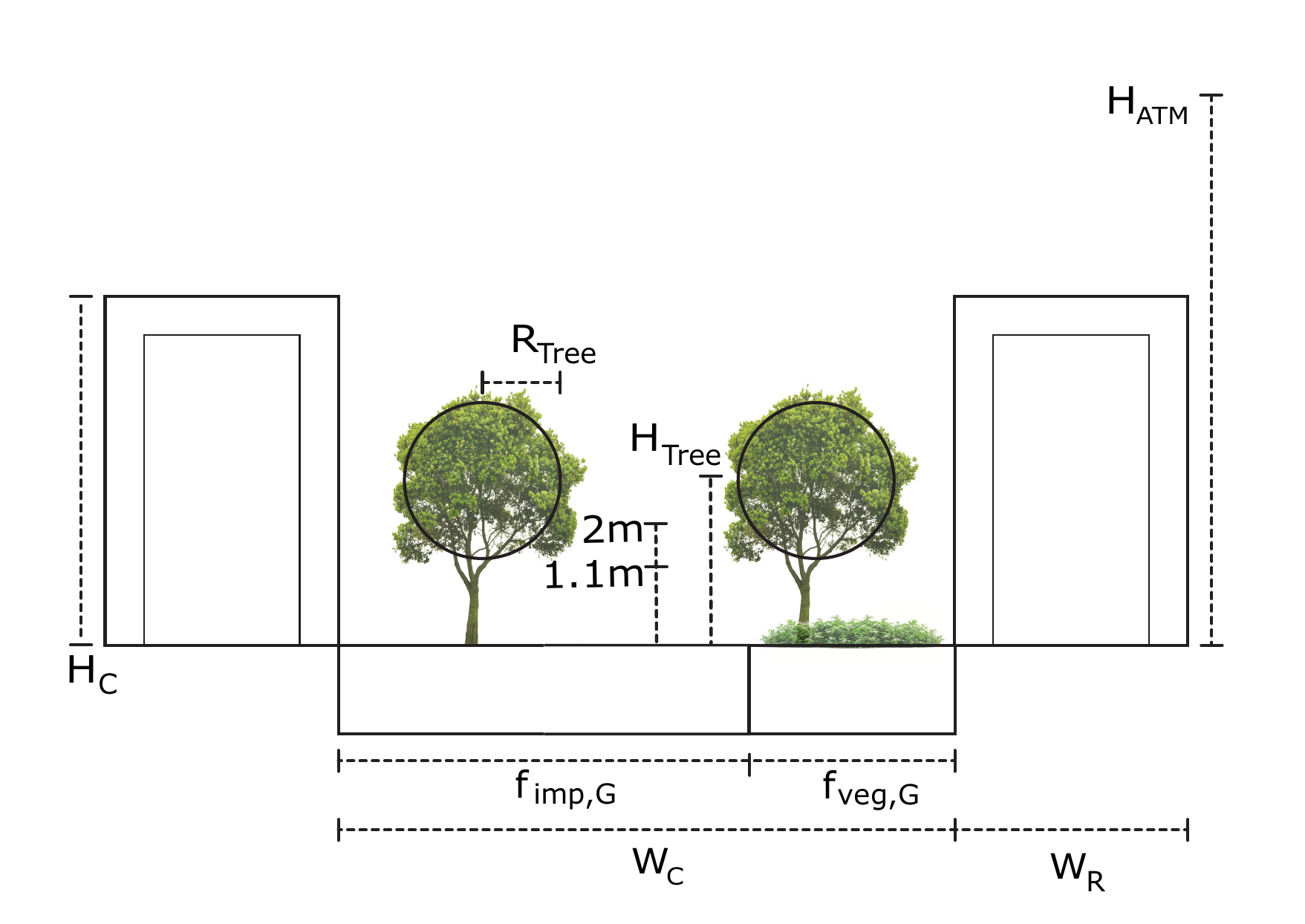}
    \caption{Schematic of the modeled urban canyon adapted from \cite{meili2020utc} with indications for modeling heights for UTCI and $T_{2\text{m}}$.}
    \label{fig:canyon}
\end{figure}

The Urban Tethys–Chloris model (UT\&C) is a mechanistic single-layer urban canopy model designed to estimate meteorological variables such as air temperature $T_\mathrm{2\text{m}}$, relative humidity $RH_\mathrm{2\text{m}}$, and thermal comfort indices within an urban canyon \citep{meili2020utc}. Meteorological forcing data $M_\mathrm{\text{atm}}(t)$ serves as input to the model function $f_\mathrm{\text{UT\&C}}$:

\begin{equation}
   M_{\mathrm{atm}}(t)=
\left[SW_{\mathrm{in}}(t),LW_{\mathrm{in}}(t),T_{\mathrm{atm}}(t),RH_{\mathrm{atm}}(t),P(t),v_{\mathrm{atm}}(t),p_{\mathrm{atm}}(t)\right].
\end{equation}
with $SW_\mathrm{\text{in}}(t)$ denoting the incoming shortwave and $LW_\mathrm{\text{in}}(t)$ incoming longwave radiation, $T_\mathrm{\text{atm}}(t)$ the atmospheric forcing temperature, $RH_\mathrm{\text{atm}}(t)$ the atmospheric relative humidity, $P(t)$ precipitation, $v_\mathrm{\text{atm}}(t)$ wind speed and $p_\mathrm{\text{atm}}(t)$ atmospheric pressure. The model returns time series of temperature $T_\mathrm{2{\text{m}}}$, relative humidity $RH_\mathrm{\text{2m}}$, and $\text{UTCI}_\mathrm{1.1m}$:
\begin{equation}\label{eq:utc_abstract}
    \left[T_\mathrm{2\text{m}}(t;\boldsymbol{\beta}), RH_\mathrm{2\text{m}}(t;\boldsymbol{\beta}), \text{UTCI}_\mathrm{1.1\text{m}}(t;\boldsymbol{\beta})\right] = f_\mathrm{\text{UT\&C}}(M_\mathrm{\text{atm}}(t);\boldsymbol{\beta})
\end{equation}
The model is implemented in MATLAB and we developed all additional functionalities with minor changes to the original code.

UT\&C simulates energy and water fluxes in an infinite two-dimensional urban canyon (Figure \ref{fig:canyon}), fully resolving the energy and water balance. Based on an integrated ecohydrological model, it accounts for urban tree cover, short ground vegetation, and green roofs, incorporating ecophysiological characteristics of different plant types as well as biochemical properties that are driving the photosynthetic processes \citep{fatichi_mechanistic_2012}. Developed in Zurich and validated for four cities in different climate zones \citep{meili2021trees}, the model determines hourly temperature $T_\mathrm{\text{2m}}$ and relative humidity $RH_\mathrm{\text{2m}}$ at a height of 2 m in the middle of the canyon, as well as $\text{UTCI}_\mathrm{1.1\text{m}}$ at a hypothetical center of a person at 1.1 m height. For meteorological forcing, data on radiation parameters, atmospheric temperature, precipitation, and relative humidity at an atmospheric reference height are required \citep{meili2020utc}. 

The infinite, two-dimensional urban canyon is characterized by an average building height $H_\mathrm{\text{c}}$ as well as roof and street width $W_\mathrm{\text{r}}$ and $W_\mathrm{\text{c}}$ \citep{meili2020utc}. The geometry is the main driver of the radiative transfer of energy by determining the fraction of shaded surfaces, along with surface parameters such as the albedo of roofs $\alpha_\mathrm{\text{r}}$ and walls $\alpha_\mathrm{\text{w}}$ that define the proportion of reflected radiation. The turbulent and conductive transport of heat and moisture between the urban structures and the atmosphere is determined by the overall urban geometry and roughness as well as material parameters. UT\&C features an integrated building energy model. Anthropogenic heat emissions are added as sensible heat sources to the canyon. In this study, neither air conditioning (AC) nor heating is applied as buildings typically do not feature AC in the current mid-European climate and heating can be neglected in summer. Instead, constant emissions of 10 W/m² are assumed for the residential area \citep{Oke_Mills_Christen_Voogt_2017}. Trees are represented as two rows along the street that are characterized by the distance to the nearest wall $D_\mathrm{\text{tree}}$ and a uniform crown radius $R_\mathrm{\text{tree}}$ and height $H_\mathrm{\text{tree}}$. The ground vegetation $f_\mathrm{\text{veg,G}}$ is characterized by the fraction of vegetated surfaces compared to bare $f_\mathrm{\text{bare}}$ and impervious surfaces $f_\mathrm{\text{imp,G}}$. Vegetation processes are further governed by the leaf area index and biochemical properties. For a detailed description of the incorporated processes, see \cite{meili_supplement_2020}. In order to validate the model, in this study, $T_\mathrm{2\text{m}}$ and $RH_\mathrm{2\text{m}}$ were simulated for the summer study period and compared with observational data from the two streets in Braunschweig.

\subsection{Multi-objective Bayesian Optimization Framework}\label{sec:pareto}
\begin{figure}[htbp]
\centering
\includegraphics[width=0.95\linewidth]{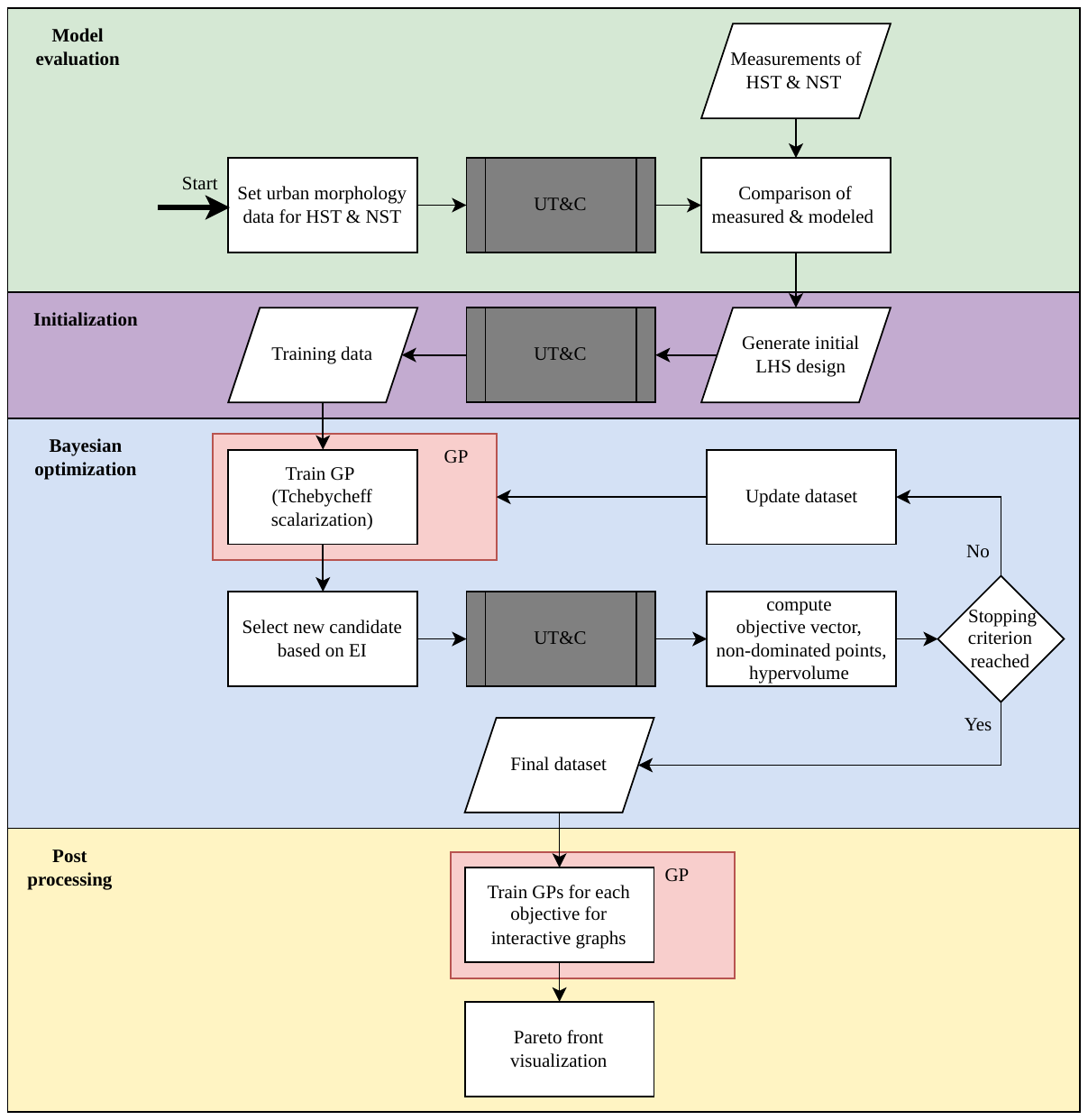}
\caption{Workflow of the proposed optimization framework.}
\label{fig:Flowchart}
\end{figure}

This section presents the framework that connects the UT\&C and the thermal metrics introduced above to the multi-objective optimization procedure. Figure~\ref{fig:Flowchart} summarizes the overall workflow, from initial parameter space sampling to Tchebycheff scalarization and Gaussian Process (GP) surrogate modeling in the iterative BO loop. The following subsections describe these components in detail.

\paragraph{Parameter space and design variables}

The full input space of the UT\&C model is too high-dimensional for exhaustive exploration. Therefore, a reduced parameter space is considered based on the sensitivity analysis of \cite{liStudyParameterCalibration2026}. The selected parameters represent the urban characteristics that were found to exert the strongest influence on near-surface thermal conditions while remaining amenable to urban climate mitigation strategies.
The optimization is performed over the eight-dimensional parameter vector
\begin{equation}
\label{eq:parameters}
\boldsymbol{\beta}
=
\left[
H_{\mathrm{c}},\,
W_{\mathrm{c}},\,
W_{\mathrm{r}},\,
f_{\mathrm{veg,G}},\,
\alpha_{\mathrm{tree}},\,
\alpha_{\mathrm{w}},\,
\lambda_{\mathrm{w}},\,
c_{\mathrm{v,s,w}}
\right]^T  \in \mathbb{R}^{8}.
\end{equation}
The corresponding parameter ranges are summarized in Table~\ref{tab:param_ranges}. They are chosen to represent realistic urban configurations observed in European cities while remaining within the geometric constraints of the UT\&C model and below the atmospheric forcing height.
\begin{table}
\centering
\caption{Parameter ranges used in the optimization.}
\label{tab:param_ranges}
\begin{tabular}{lccc}
\hline
\textbf{Parameter} & \textbf{Symbol} & \textbf{Lower bound} & \textbf{Upper bound} \\
\hline
\multicolumn{4}{l}{\textit{Geometric parameters}} \\
\hline
Canyon height [m] & $H_\mathrm{\text{c}}$ & 15.0 & 24.9 \\
Canyon width [m] & $W_\mathrm{\text{c}}$ & 15.0 & 30.0 \\
Roof width [m] & $W_\mathrm{\text{r}}$ & 8.0 & 25.0 \\
\hline
\multicolumn{4}{l}{\textit{Mitigation parameters}} \\
\hline
Ground vegetation fraction [-] & $f_\mathrm{\text{veg,G}}$ & 0.0 & 1.0 \\
Intrinsic tree radius scaling [-] & $\alpha_\mathrm{\text{tree}}$ & 0.02 & 0.24 \\
Albedo (wall) [-] & $\alpha_\mathrm{\text{w}}$ & 0.25 & 0.90 \\
Thermal conductivity (wall) [W\,m$^{-1}$\,K$^{-1}$] & $\lambda_\mathrm{\text{w}}$ & 0.1 & 3.8 \\
Vol. heat capacity (wall) [J\,m$^{-3}$\,K$^{-1}$] & $c_\mathrm{\text{v,s,w}}$ & $0.1\times10^6$ & $3.2\times10^6$ \\
\hline
\end{tabular}
\end{table}
The parameter set comprises urban geometry, vegetation characteristics, and material properties that govern the canyon energy balance:
\begin{itemize}
    \item The geometric parameters, building height $H_\mathrm{\text{c}}$, canyon width $W_\mathrm{\text{c}}$, and roof width $W_\mathrm{\text{r}}$, characterize the urban morphology and determine the radiative and aerodynamic environment within the street canyon. Although these quantities are difficult to modify in existing urban environments, their inclusion allows the framework to assess how the effectiveness of mitigation measures depends on the surrounding urban form.
    \item Vegetation is represented through the ground vegetation fraction $f_\mathrm{\text{veg,G}}$ and the tree crown scaling parameter $\alpha_\mathrm{\text{tree}}$. The vegetation fraction ranges from fully impervious to fully vegetated ground surfaces, whereas rooftop vegetation is not considered. Tree geometry is parameterized through 
    \begin{equation}
        \label{eq:Tree}
        R_\mathrm{\text{tree}} = \alpha_\mathrm{\text{tree}} *W_\mathrm{\text{c}},
    \end{equation}
    where $R_\mathrm{\text{tree}}$ denotes the tree crown radius and $\alpha_\mathrm{\text{tree}}$ controls its size relative to the canyon width. This formulation ensures geometrically consistent tree configurations across the entire design space. Tree height is fixed at $7.5 \, \mathrm{m}$, so that the tree geometry remains compatible with all canyon geometries considered. Since the UT\&C model requires $R_\mathrm{\text{tree}}$ as an input, the corresponding radius is calculated from $\alpha_\mathrm{\text{tree}}$ and $W_\mathrm{\text{c}}$ for each model evaluation.

    \item The remaining parameters describe thermophysical surface properties relevant to urban heat mitigation. Wall albedo $\alpha_\mathrm{w}$ controls the reflection of incoming shortwave radiation, while thermal conductivity (wall) $\lambda_\mathrm{w}$ and volumetric heat capacity (wall) $c_{v,s,w}$ determine heat storage and heat transfer within the building envelope. The selected ranges cover typical building materials and insulation standards encountered in European cities \citep{johra_2021}.

\end{itemize}
To maintain consistency with the observational reference site, canyon orientation is fixed to the HST configuration (Table~\ref{tab:utc_eval_params}). Nevertheless, the proposed framework is not restricted to this setting and can readily be extended to alternative orientations or additional mitigation measures. Rather than prescribing discrete mitigation strategies, the framework explores a continuous urban design space within which such interventions operate. This enables the identification of parameter combinations that improve urban thermal conditions while accounting for interactions between urban morphology, vegetation, and material properties.

\paragraph{Pareto optimality}

The objective vector $\mathbf{y}$ associated with a parameter vector $\boldsymbol{\beta}$ is constructed from the corresponding UT\&C output time series by evaluating the three thermal metrics defined in Eqs.~\eqref{eq:peak}--\eqref{eq:UTCIpeak}:
\begin{equation}\label{eq:objective_vector}
    \mathbf{y}(\boldsymbol{\beta}) = \left[
        T_{\mathrm{peak}}(\boldsymbol{\beta}),\
        T_{\mathrm{night}}(\boldsymbol{\beta}),\
        \mathrm{UTCI}_{\mathrm{peak}}(\boldsymbol{\beta})
    \right]^T \in \mathbb{R}^{3}.
\end{equation}

Since improving one objective often comes at the expense of another, no single objective vector simultaneously minimizes all three objectives. Instead, objective vectors are compared using Pareto dominance. An objective vector $\mathbf{y}(\boldsymbol{\beta}_i)$ is said to be \emph{dominated} by another objective vector $\mathbf{y}(\boldsymbol{\beta}_j)$ if

\begin{equation}
\begin{aligned}
y_m(\boldsymbol{\beta}_j) &\leq y_m(\boldsymbol{\beta}_i)
&& \forall\, m \in \{1,2,3\}, \\
y_m(\boldsymbol{\beta}_j) &< y_m(\boldsymbol{\beta}_i)
&& \text{for at least one } m.
\end{aligned}
\end{equation}

An objective vector is \emph{Pareto optimal} if it is not dominated by any other objective vector. The global set of all Pareto-optimal objective vectors is referred to as the \emph{Pareto front}, which represents the set of achievable trade-offs between the three objectives \citep{collette_multiobjective_2004}.

In practice, the true Pareto front is generally unknown, particularly for computationally expensive simulation models, since determining it would require an exhaustive exploration of the design space. Consequently, the non-dominated solutions identified during the optimization process represent only a discrete approximation of the true Pareto front rather than the complete set of globally Pareto-optimal solutions. The objective is therefore to iteratively improve this approximation to efficiently capture trade-offs between the three thermal metrics while limiting the number of computationally expensive UT\&C model evaluations \citep{macasieb_probabilistic_2024}.

\paragraph{Augmented Tchebycheff scalarization}

In the present implementation, the optimization is performed on a scalar objective function to enable the application of single-objective Bayesian optimization while preserving the trade-offs between the individual objectives. The objective vector defined in Eq.~\eqref{eq:objective_vector} is therefore transformed into a scalarized objective function using the augmented Tchebycheff scalarization for BO proposed by \citet{knowlesParEGOHybridAlgorithm2006}:

\begin{equation}
\label{eq:Tcheby}
J(\boldsymbol{\beta})
=
\max_{m=1,\ldots,3}
\left\{
\lambda_m
\tilde{y}_m(\boldsymbol{\beta})
\right\}
+
\rho
\sum_{m=1}^{3}
\lambda_m
\tilde{y}_m(\boldsymbol{\beta}),
\end{equation}

where $\boldsymbol{\lambda} = (\lambda_1, \lambda_2, \lambda_3)^T$ is a weight vector on the unit simplex, i.e. $\sum_{m=1}^{3} \lambda_m = 1$, obtained by mapping a two-dimensional scrambled Sobol' sequence (Owen scrambling) onto the simplex via
\begin{equation}
    \lambda_1 = 1 - \sqrt{u_1}, \qquad
    \lambda_2 = \sqrt{u_1}\,(1 - u_2), \qquad
    \lambda_3 = \sqrt{u_1}\,u_2,
\end{equation}
with $(u_1, u_2) \in [0,1]^2$ drawn from the scrambled Sobol' sequence, ensuring low-discrepancy coverage of the weight space. $\rho$ is a small positive augmentation parameter, here set to $\rho = 0.05$.

Since the three objectives differ in scale, they are normalized using the current minimum and maximum values observed for each objective,

\begin{equation}
\label{eq:span}
\tilde{y}_m(\boldsymbol{\beta})
=
\frac{y_m(\boldsymbol{\beta})-y_m^{\min}}
{y_m^{\max}-y_m^{\min}},
\qquad
m=1,\ldots,3.
\end{equation}

The augmented Tchebycheff scalarization is used to explore different trade-offs between the objectives by varying the weighting vector $\boldsymbol{\lambda}_k$. The additional augmentation term helps distinguish weakly Pareto-optimal solutions from Pareto-optimal solutions by favoring solutions that improve multiple objectives simultaneously.

\paragraph{Bayesian optimization procedure}

BO is employed to efficiently explore the parameter space while limiting the number of computationally expensive UT\&C model evaluations. The optimization is initialized using Latin Hypercube Sampling (LHS) of $N=50$ parameter vectors within the bounds defined in Table~\ref{tab:param_ranges}, yielding the initial dataset

\begin{equation}
\mathcal{D}
=
\left\{
\left(
\boldsymbol{\beta}_i,
\mathbf{y}(\boldsymbol{\beta}_i)
\right)
\right\}_{i=1}^{N}.
\end{equation}

Given the dimensionality of the design space, 50 samples cannot provide an exhaustive representation of all feasible parameter combinations or the corresponding objective space. Instead, the initial LHS design serves exclusively as a space-filling initialization for the Gaussian Process model, enabling Bayesian optimization to iteratively improve the surrogate in the most informative regions of the parameter space. \citep{jones_efficient_1998}

At each BO iteration $k$, a weight vector $\boldsymbol{\lambda}_k$ is drawn and Eqs.~\eqref{eq:Tcheby} and~\eqref{eq:span} are evaluated with $\boldsymbol{\lambda}=\boldsymbol{\lambda}_k$ and the objective ranges currently observed in $\mathcal{D}$, giving $\tilde{y}_{m,k}(\boldsymbol{\beta})$ and the iteration-specific scalarized objective $J_k(\boldsymbol{\beta})$. The multi-objective optimization problem is thereby transformed into the scalar optimization problem

\begin{equation}
\label{eq:scalar_optimization}
\boldsymbol{\beta}_k^{*}
=
\arg\min_{\boldsymbol{\beta}\in\mathcal{B}}
J_k(\boldsymbol{\beta}),
\end{equation}

\noindent where $\mathcal{B}$ denotes the feasible parameter space defined by the parameter bounds (Table~\ref{tab:param_ranges}).

For this purpose, a GP is trained at every iteration to approximate the current scalarized objective function $J_k(\boldsymbol{\beta})$. The GP prior is given by

\begin{equation}
J_k(\boldsymbol{\beta})
\sim
\mathcal{GP}
\bigl(
m(\boldsymbol{\beta}),
\kappa(\boldsymbol{\beta},\boldsymbol{\beta}')
\bigr),
\end{equation}

where $m(\boldsymbol{\beta})$ denotes the mean function and $\kappa(\boldsymbol{\beta},\boldsymbol{\beta}')$ the covariance kernel. An automatic relevance determination squared exponential kernel is employed with standardized predictors. The kernel hyperparameters, including the observation noise variance, are estimated by maximizing the marginal log-likelihood \citep{rasmussenGaussianProcessesMachine2008}.

Because a new weight vector $\boldsymbol{\lambda}_k$ is sampled at every iteration, successive iterations emphasize different trade-offs between the three objectives and the approximation of the Pareto front is progressively refined. At iteration $k=1,2,\ldots$, the GP is trained on the current dataset $\mathcal{D}$ and yields the predictive distribution

\begin{equation}
\label{eq:posterior}
p\bigl(J_k(\boldsymbol{\beta})\mid\mathcal{D}\bigr)
=
\mathcal{N}
\bigl(
\mu_k(\boldsymbol{\beta}),
\sigma_k^2(\boldsymbol{\beta})
\bigr),
\end{equation}

where $\mu_k(\boldsymbol{\beta})$ and $\sigma_k^2(\boldsymbol{\beta})$ denote the predictive mean and variance at iteration $k$, respectively.

The next candidate to be evaluated by the UT\&C model is selected by maximizing the Expected Improvement (EI) acquisition function \citep{jones_efficient_1998},

\begin{equation}
\label{eq:EI}
\mathrm{EI}_k(\boldsymbol{\beta})
=
\mathbb{E}
\left[
\max
\left(
J_{k,\min}
-
J_k(\boldsymbol{\beta}),
0
\right)
\right],
\end{equation}

where $J_{k,\min}$ denotes the smallest scalarized objective value observed up to iteration $k$ and $\mathbb{E}[\cdot]$ denotes the expectation with respect to the GP posterior. Since the posterior distribution in Eq.~\eqref{eq:posterior} is Gaussian, the expectation in Eq.~\eqref{eq:EI} admits the closed-form expression

\begin{equation}
\mathrm{EI}_k(\boldsymbol{\beta})
=
\left(
J_{k,\min}
-
\mu_k(\boldsymbol{\beta})
\right)
\Phi(z_k)
+
\sigma_k(\boldsymbol{\beta})
\phi(z_k),
\end{equation}

with

\begin{equation}
z_k
=
\frac{
J_{k,\min}
-
\mu_k(\boldsymbol{\beta})
}
{
\sigma_k(\boldsymbol{\beta})
},
\end{equation}

where $\Phi(\cdot)$ and $\phi(\cdot)$ denote the cumulative distribution function and probability density function of the standard normal distribution, respectively.

At each BO iteration $k$, a candidate pool of $N_c=50\,000$ parameter vectors is generated using a Sobol sequence. $\mathrm{EI}_k$ is evaluated for every candidate using the current GP surrogate, and the ten highest-ranked candidates are used as initial guesses for a gradient-based optimization (\texttt{fmincon}) to maximize $\mathrm{EI}_k$. The resulting parameter vector $\boldsymbol{\beta}_{N+k}$ is evaluated using the UT\&C model, and the corresponding objective vector is appended to the dataset

\begin{equation}
\label{eq:dataset_update}
\mathcal{D}
\leftarrow
\mathcal{D}
\cup
\left\{
\left(
\boldsymbol{\beta}_{N+k},
\mathbf{y}(\boldsymbol{\beta}_{N+k})
\right)
\right\}.
\end{equation}

After each BO iteration $k$, the hypervolume $HV_k$ enclosed by the current non-dominated objective vectors is computed. These objective vectors constitute the current approximation of the Pareto front at iteration $k$. The hypervolume measures the volume of the objective space dominated by this approximation and bounded by a fixed reference point $\mathbf{y}^{\mathrm{ref}}$, defined component-wise from the worst objective vectors observed in the initial dataset (the $N=50$ LHS samples) and scaled by a factor of $1.2$ to ensure the reference remains dominated by all objective vectors encountered during the optimization.

The optimization terminates when the relative increase in hypervolume,
\begin{equation}
\label{eq:hv_stop}
\Delta HV_k = \frac{HV_k - HV_{k-w}}{HV_{k-w}},
\end{equation}
over the previous $w=20$ iterations falls below $\varepsilon_{HV}=0.1\%$. Additionally, a minimum number of $k_{\min}=50$ and a maximum of $k_{\max}=500$ BO iterations is imposed.

The complete workflow is summarized in Algorithm~\ref{alg:framework}.

\begin{algorithm}[htbp]
\caption{Multi-Objective Bayesian Optimization using Augmented Tchebycheff Scalarization}
\label{alg:framework}
\begin{algorithmic}[1]

\Require Initial design size $N=50$, candidate pool size $N_c=50\,000$, reference point $\mathbf{y}^{\mathrm{ref}}$, hypervolume tolerance $\varepsilon_{HV}=0.1\%$, hypervolume window $w=20$, minimum BO iterations $k_{\min}=50$, maximum BO iterations $k_{\max}=500$
\Ensure Approximation of the Pareto front

\State Generate an initial LHS design with $N$ parameter vectors $\boldsymbol{\beta}_i$

\For{$i=1,\ldots,N$}
    \State Evaluate the UT\&C model
    \State Compute objective vector
    $
    \mathbf{y}(\boldsymbol{\beta}_i)
    =
    \left[
    T_{\mathrm{peak}},
    T_{\mathrm{night}},
    \mathrm{UTCI}_{\mathrm{peak}}
    \right]^\top$
\EndFor

\State Construct the initial dataset
$
\mathcal{D}
=
\left\{
\left(
\boldsymbol{\beta}_i,
\mathbf{y}(\boldsymbol{\beta}_i)
\right)
\right\}_{i=1}^{N}
$

\State Set the reference point $\mathbf{y}^{\mathrm{ref}} = 1.2 \times \bigl(\text{component-wise worst values in } \mathcal{D}\bigr)$

\For{$k=1,\ldots,k_{\max}$}

    \State Draw a random weight vector $\boldsymbol{\lambda}_k$ from the unit simplex

    \State Normalize each objective
    \[
    \tilde y_{m,k}(\boldsymbol{\beta})
    =
    \frac{
    y_{m,k}(\boldsymbol{\beta})-y_{m,k}^{\min}
    }{
    y_{m,k}^{\max}-y_{m,k}^{\min}
    }
    \]

    \State Compute the scalarized objective
    \[
    J_k(\boldsymbol{\beta})
    =
    \max_m
    \left\{
    \lambda_{m,k}\,\tilde y_{m,k}(\boldsymbol{\beta})
    \right\}
    +
    \rho
    \sum_m
    \lambda_{m,k}\,\tilde y_{m,k}(\boldsymbol{\beta}),
    \qquad
    m=1,\ldots,3
    \]

    \State Train the GP surrogate on the scalarized dataset $\{(\boldsymbol{\beta}_i,J_k(\boldsymbol{\beta}_i))\}_{i=1}^{N+k-1}$

    \State Construct the acquisition function $\mathrm{EI}_k(\boldsymbol{\beta})$

    \State Generate $N_c$ Sobol candidate points

    \State Evaluate $\mathrm{EI}_k(\boldsymbol{\beta})$ for all $N_c$ candidates using the GP

    \State Select the ten candidates with the largest $\mathrm{EI}_k$ values

    \State Maximize $\mathrm{EI}_k(\boldsymbol{\beta})$ using \texttt{fmincon} with these ten initial guesses

    \State Obtain the next evaluation point
    $
    \boldsymbol{\beta}_{N+k}
    =
    \arg\max_{\boldsymbol{\beta}}
    \mathrm{EI}_k(\boldsymbol{\beta})
    $

    \State Evaluate the UT\&C model at $\boldsymbol{\beta}_{N+k}$

    \State Compute
    $\mathbf{y}(\boldsymbol{\beta}_{N+k})$

    \State Update
    $
    \mathcal{D}
    \leftarrow
    \mathcal{D}
    \cup
    \left\{
    \left(
    \boldsymbol{\beta}_{N+k},
    \mathbf{y}(\boldsymbol{\beta}_{N+k})
    \right)
    \right\}$

    \State Determine the current non-dominated objective vectors $\{\mathbf{y}(\boldsymbol{\beta}_i)\}_{i=1}^{N+k}$
    \State Compute the hypervolume $HV_k$ with respect to $\mathbf{y}^{\mathrm{ref}}$
    \If{$k \ge \max(w,k_{\min})$}
        \State Compute
        \[
        \Delta HV_k
        =
        \frac{HV_k-HV_{k-w}}
        {HV_{k-w}}
        \]

        \If{$\Delta HV_k<\varepsilon_{HV}$}
            \State \textbf{break}
        \EndIf
    \EndIf

\EndFor

\end{algorithmic}
\end{algorithm}

\paragraph{Post-processing and visualization}
\label{par:Post}
Following convergence of the BO procedure, the resulting dataset of evaluated parameter vectors $\boldsymbol{\beta}$ and corresponding thermal metrics $\mathbf{y}(\boldsymbol{\beta})$, is used to train three separate GP surrogates, each mapping the parameter vector $\boldsymbol{\beta}$ onto one scalar objective $y_m(\boldsymbol{\beta})$, for $m=1,2,3$, corresponding to $T_{\mathrm{peak}}$, $T_{\mathrm{night}}$, and $\mathrm{UTCI}_{\mathrm{peak}}$, respectively. The GP prior therefore reads as

\begin{equation}
y_m(\boldsymbol{\beta})
\sim
\mathcal{GP}
\bigl(
m(\boldsymbol{\beta}),
\kappa(\boldsymbol{\beta},\boldsymbol{\beta}')
\bigr),
\end{equation}
for each $m$. 

Each surrogate is trained independently, with its own kernel hyperparameters estimated separately by maximizing the respective marginal log-likelihood; no information is shared between the three GPs. An 85\%/15\% train--test split is used to assess the predictive performance of each surrogate. These  surrogates are then retrained on the full dataset and used for post-processing, without requiring further evaluations of the computationally expensive UT\&C model.

First, the GPs are used to generate a dense approximation of the Pareto front. To this end, the augmented Tchebycheff scalarization described above is repeatedly applied to the mean functions of the 3 GP predictive posteriors, using randomly sampled weight vectors. For each sampled weight vector, the resulting scalarized objective is minimized over the parameter space using a gradient-based optimizer (\texttt{fmincon}) with multiple randomized starting points to mitigate convergence to local optima. This differs fundamentally from the BO loop described above: rather than selecting the next evaluation point via an acquisition function that balances exploration and exploitation under GP uncertainty, the predictive means are treated as a sufficiently accurate surrogate of the true objective space, and the scalarized cost is minimized directly for each sampled weight vector, without updating the dataset $\mathcal{D}$. This enables dense sampling of the Pareto front approximation and provides a compact representation of the trade-offs between $T_{\mathrm{peak}}$, $T_{\mathrm{night}}$, and $\mathrm{UTCI}_{\mathrm{peak}}$ within the considered design space.

Second, the GPs are used for a global sensitivity analysis. Total-order Sobol indices are estimated from the GP predictions, allowing the contribution of each design parameter to the variance of the three thermal metrics to be quantified \citep{saltelli_variance_2010}.

To formalize this step, consider again the parameter vector $\boldsymbol{\beta}$ defined in \eqref{eq:parameters}. While $\boldsymbol{\beta}_i$ and $\boldsymbol{\beta}_{N+k}$ (as used above) denote individual \emph{sampled} parameter vectors, the sensitivity analysis instead considers variation along individual \emph{components} of $\boldsymbol{\beta}$. We denote by $\beta_l$, $l = 1, \dots, 8$, the $l$-th component of $\boldsymbol{\beta}$, corresponding to $H_{\mathrm{c}},\,
W_{\mathrm{c}},\,
W_{\mathrm{r}},\,
f_{\mathrm{veg,G}},\,
\alpha_{\mathrm{tree}},\,
\alpha_{\mathrm{w}},\,
\lambda_{\mathrm{w}},\,
c_{\mathrm{v,s,w}}$, respectively.

The total-order Sobol index quantifies the contribution of component $\beta_l$ to the output variance, including all interactions with other components, and is given by
\begin{equation}
    S_{T_l}^{(m)} =
    \frac{
    \mathbb{E}_{\boldsymbol{\beta}_{\sim l}}
    \left[
    \operatorname{Var}_{\beta_l}
    \left(
    y_m(\boldsymbol{\beta})
    \mid
    \boldsymbol{\beta}_{\sim l}
    \right)
    \right]
    }{
    \operatorname{Var}
    \left(
    y_m(\boldsymbol{\beta})
    \right)
    },
    \qquad m \in \{1,2,3\}.
    \label{eq:total_sobol}
\end{equation}

\noindent
where $\boldsymbol{\beta}_{\sim l}$ denotes the parameter vector with $\beta_l$ excluded, and $y_m(\boldsymbol{\beta})$ denotes the $m$-th component of the objective vector $\mathbf{y}(\boldsymbol{\beta})$ defined in Eq.~\eqref{eq:objective_vector}, with $m \in \{1,2,3\}$ corresponding to $T_{\mathrm{peak}}$, $T_{\mathrm{night}}$, and $\mathrm{UTCI}_{\mathrm{peak}}$, respectively.

Finally, the three GPs are incorporated into the \emph{Urban Climate Design Explorer}, an interactive visualization tool that enables real-time exploration of parameter interactions and trade-offs between the three thermal metrics. By providing instantaneous predictions throughout the design space, the tool facilitates the interpretation of the optimization results and supports the assessment of urban climate mitigation strategies. The \emph{Urban Climate Design Explorer} is made publicly available through a GitHub repository to promote reproducibility and facilitate its application in future studies \citep{walter_respace_interaction_2026}.
 
\newpage
\section{Results and discussion}
This chapter first evaluates the performance of the UT\&C model for the two study streets in~\ref{sec:utc_results}. Second, the results of the optimization are analyzed from a methodological perspective, including the characteristics of the Pareto front, and the robustness of the proposed framework in Section~\ref{sec:optimization_results}. Then, the optimization results are translated into urban heat mitigation strategies using both the Pareto analysis and the interactive \textit{Urban Climate Design Explorer}, which enables systematic exploration of the influence of individual design parameters on the thermal metrics in Section~\ref{sec:evaluation}. Finally, limitations of the presented methodology are critically discussed in Section~\ref{sec:limitations}.

\subsection{UT\&C model evaluation within different street configurations}\label{sec:utc_results}

In this section, we verify the ability of the UT\&C model to reproduce the influence of contrasting urban geometry and vegetation characteristics on $T_\mathrm{2\text{m}}$ and $RH_\mathrm{2\text{m}}$. $T_\mathrm{\text{2m}}$ and $RH_\mathrm{\text{2m}}$ were simulated for the study period and compared with observational data collected at the two street canyon sites in Braunschweig, using the coefficient of determination ($R^2$), mean error, and root mean square error (RMSE). UT\&C provides air temperature and RH at a height of 2 m, whereas field measurements were conducted at a height of 3 m, as for safety reasons, this is the standard height used in the Braunschweig measuring network \citep{grunwald_mapping_2019}.
Global error metrics are reported in Table~\ref{tab:utc_eval_metrics}. Generally, the model evaluation for the two urban street canyons in Braunschweig shows that the model is capable of accurately reproducing the impact of different street geometry and vegetation characteristics within the same mesoscale climate, in particular for air temperature. Further details regarding the model performance for $T_\mathrm{2\text{m}}$ and $RH_\mathrm{2\text{m}}$ can be found in Appendix~\ref{sec:verification}. Appendix~\ref{sec:verification} describes the dimension of the differences throughout the day, places the findings in the context of previous literature, and discusses possible explanations for the observed differences.

\begin{table}[htbp]
\centering
\caption{Mean error and RMSE statistics for temperature and relative humidity under HST and NST configurations. Daytime was defined as the time frame between 6:00 and 21:00.} 
\label{tab:utc_eval_metrics}
\begin{tabular}{llcccccc}
\hline
\multirow{2}{*}{Street} & \multirow{2}{*}{Metric} &
\multicolumn{3}{c}{Temperature [$^\circ$C]} &
\multicolumn{3}{c}{Relative humidity [\%]} \\
\cline{3-8}
 & & Total & Day & Night & Total & Day & Night \\
\hline
\multirow{2}{*}{HST}
& Mean error
& -0.14 & -0.01 & -0.40
& 2.66 & 3.16 & 1.66 \\
& RMSE
& 0.78 & 0.80 & 0.73
& 5.81 & 6.16 & 5.02 \\
\hline
\multirow{2}{*}{NST}
& Mean error
& -0.35 & -0.28 & -0.47
& 0.10 & -0.24 & 0.79 \\
& RMSE
& 0.79 & 0.80 & 0.76
& 3.95 & 4.17 & 3.45 \\
\hline
\end{tabular}
\end{table}

\subsection{Interpretation and discussion of optimization results}\label{sec:optimization_results}

\begin{figure}[htbp]
    \centering
    \includegraphics[width=1\linewidth]{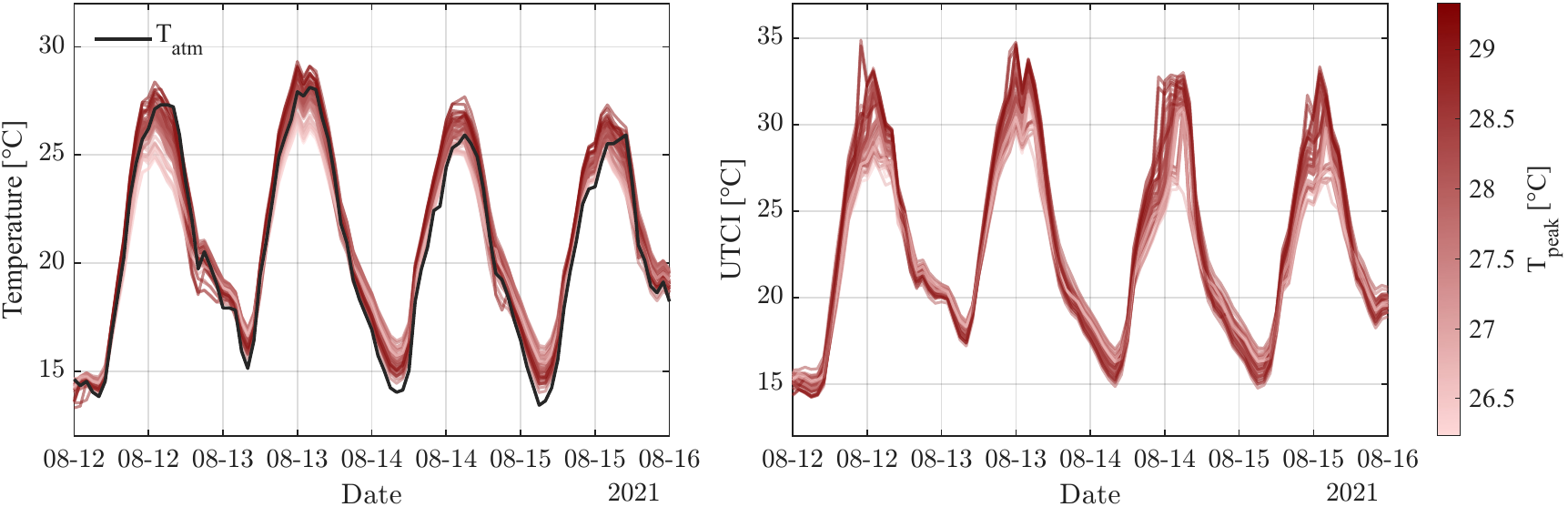}
    \caption{Temperature $T_\mathrm{2m}$ (left) and $\mathrm{UTCI}_{1.1\mathrm{m}}$ (right) time series for the $N=50$ initial LHS. Runs are colored by peak temperature $T_\mathrm{peak}$ ranging from light pink (cool) to deep red (hot). The atmospheric forcing temperature $T_\mathrm{atm}$ is shown in black for reference.}
    \label{fig:timeseries_all}
\end{figure}

In the following, we examine the outcome of the multi-objective BO, where we optimized urban morphology and vegetation parameters jointly to minimize daytime heat stress ($T_\mathrm{peak}$ and $\mathrm{UTCI}_\mathrm{peak}$) while promoting nocturnal cooling ($T_\mathrm{night}$). We first assess the optimization outcome itself, i.e. convergence behavior of the Pareto front by evaluating the hypervolume that spans the Pareto front, before examining which parameters drive the resulting trade-off between daytime heat mitigation and nocturnal cooling in Section~\ref{sec:evaluation}.

\paragraph{LHS Initialization} Figure~\ref{fig:timeseries_all} shows the simulated $T_\mathrm{\text{2m}}$ and $\text{UTCI}_\mathrm{1.1m}$ time series for the $50$ initially evaluated configurations. A pronounced spread is visible during daytime hours, particularly around midday, indicating that the varied parameters strongly influence daytime thermal conditions. In contrast, nighttime temperature variability is substantially smaller. The atmospheric forcing temperature lies within the upper portion of the daytime envelope but remains below the simulated nighttime minimum temperatures, indicating persistent heat storage and release within the urban canyon that limits nocturnal cooling. The $\text{UTCI}_\mathrm{1.1m}$ time series exhibit an even larger daytime spread than $T_\text{2m}$, reflecting the strong influence of radiation-related parameters on thermal comfort, while nighttime variability in $\text{UTCI}_{1.1\mathrm{m}}$ remains comparatively small.

The time series suggest that configurations associated with lower daytime temperatures tend to occupy the higher range of nighttime temperatures, directly indicating a trade-off between daytime heat mitigation and nocturnal cooling. In addition, isolated trajectories exhibit elevated temperatures during both day and night, representing clearly unfavorable configurations. At the same time, one configuration appears to achieve both comparatively low daytime and nighttime temperatures, suggesting that promising regions of the design space may exist. The conflict between objectives, together with the pronounced variability in the sampled responses, motivates a multi-objective optimization approach capable of resolving the trade-off explicitly.

\paragraph{Convergence and robustness of the optimization}

To assess the robustness of the proposed framework, the complete optimization procedure was repeated ten times using distinct, reproducible random seeds. These seeds control the quasi-randomness of the Sobol sequences used both for the Tchebycheff weight-vector generation and for the $N_c=50{,}000$ candidate points sampled at each BO iteration, resulting in different stochastic realizations of the optimization process.

Across the ten runs, the number of BO iterations required to satisfy the stopping criterion (defined as a relative hypervolume improvement below $\varepsilon_{HV}=0.1\%$ over $w=20$ consecutive iterations) ranged from $83$ to $124$, with a mean of $98.2$ iterations, indicating moderate variability in convergence speed due to the stochastic sampling strategy. In contrast, the quality of the obtained non-dominated sets remained highly consistent. The final hypervolume ranged from $884.59$ to $923.52$, with a mean of $\mu=911.77$, a standard deviation of $\sigma=12.02$, and a coefficient of variation of $\frac{\sigma}{\mu}=1.32\%$. The number of non-dominated objective vectors varied between $25$ and $38$, with a mean of $31.6$. A statistical summary of all optimization runs is provided in Tables~\ref{tab:multi_runs} and~\ref{tab:summultiruns} in Appendix~\ref{sec:verification}. 

The low variation in hypervolume demonstrates that the proposed framework consistently produces non-dominated objective vectors of comparable quality despite differences in the optimization trajectory and the number of iterations required for convergence.

\paragraph{Reference run}

Unless stated otherwise, the following analyses refer to Run~8, which satisfied the stopping criterion after $94$ BO iterations, corresponding to a total of $144$ UT\&C model evaluations ($50$ initial LHS samples and $94$ BO iterations). The resulting non-dominated set comprises $33$ objective vectors and spans a hypervolume of $914.97$. These values are close to the corresponding means across all ten optimization runs, making Run~8 a representative example for the subsequent analyses.

\begin{figure}[htbp]
    \centering
    \includegraphics[width=1\linewidth]{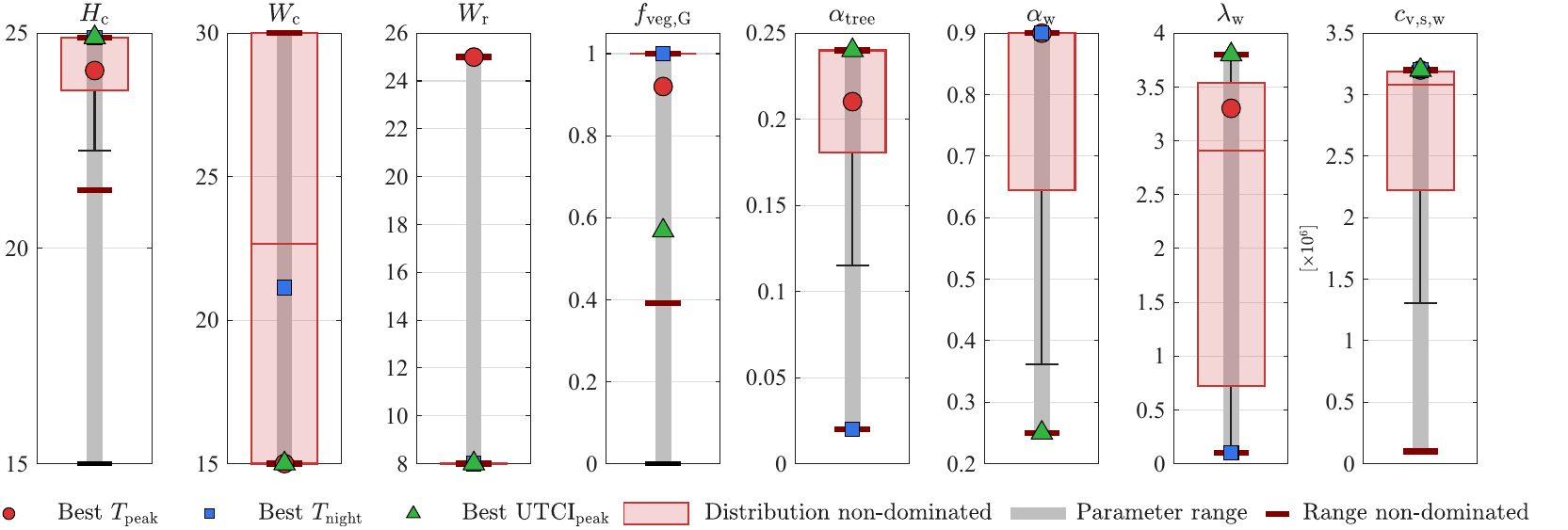}
    \caption{Design space analysis of the non-dominated set for Run 8. For each parameter, the design space, the distribution and range of the set of the 33 non-dominated parameters as well as the optimal parameter per thermal metric is shown, respectively.}
    \label{fig:HistoRange}
\end{figure}

Figure~\ref{fig:HistoRange} compares the parameter space of the non-dominated set against the overall explored design space. For the majority of parameters, the non-dominated vectors span a significant fraction of the admissible range, suggesting that multiple parameter combinations can yield comparable performance due to compensating effects between design variables. In contrast, certain parameters exhibit pronounced clustering, most notably roof width $W_\mathrm{r}$, vegetation fraction $f_\mathrm{veg,G}$, and the tree radius scaling parameter $\alpha_\mathrm{tree}$,  indicating that these variables exert a dominant influence on the objective vectors. Interestingly, the distribution of $W_\mathrm{r}$ corresponding to the non-dominated objective vectors is concentrated at the lower bound of the parameter range. In contrast, the parameter vector yielding the minimum $T_\mathrm{peak}$ is located at the upper bound, indicating a trade-off between minimizing $T_\mathrm{peak}$ and the remaining objectives.

\begin{figure}[htbp]
    \centering
    \includegraphics[width=0.99\linewidth]{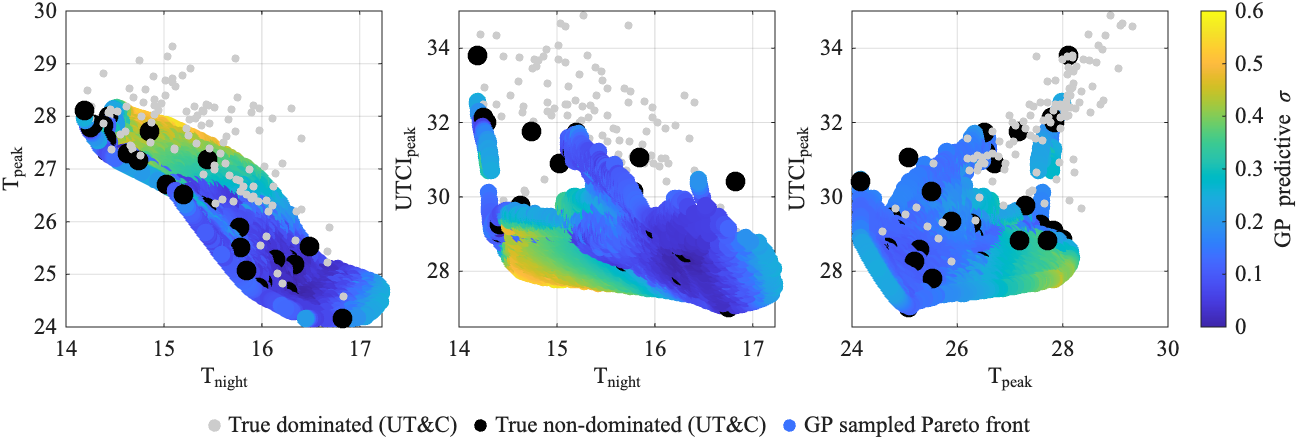}
    \caption{Pairwise projections of the three-dimensional objective space ($T_{night}$, $T_{peak}$, $UTCI_{peak}$). Black markers correspond to the 33 Pareto-optimal solutions identified by the BO framework, while gray markers represent evaluated parameter configurations yielding dominated objective vectors. Coloured markers denote the Pareto front predicted by the 3 surrogates and are colored according to the 3 combined GP's posterior predictive standard deviation $\sigma$, calculated as the Euclidean norm of their individual predictive standard deviations, with dark blue indicating the highest predictive certainty.}
    \label{fig:ParetoScatter}
\end{figure}

Based on the 144 samples, three GPs, one per thermal metric, were trained.
The surrogate models were initially validated using a training-test split (85\% training), after which they were retrained on the full set of 144 evaluations to maximize predictive accuracy. While the eight-dimensional design space remains sparsely sampled, BO progressively concentrated sampling in regions of high expected improvement for the augmented Tchebycheff cost function (Eq.~\ref{eq:Tcheby}). By doing so, the framework effectively targeted the thermally relevant regions of the design space.

\begin{figure}[htbp]
    \centering
    \includegraphics[width=0.9\linewidth]{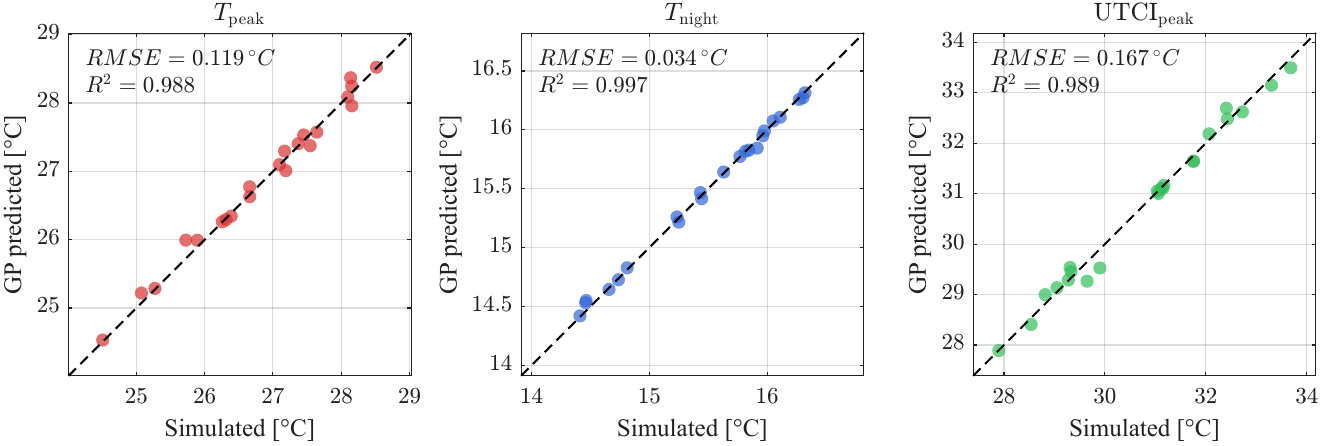}
    \caption{Validation of the GPs for $T_{\mathrm{peak}}$, $T_{\mathrm{night}}$, and $\mathrm{UTCI}_{\mathrm{peak}}$. For validation the GP models were trained on 85\% of the 144 samples; the points shown correspond to the remaining test data. The dashed line marks the line of perfect agreement between surrogate predictions and UT\&C simulation results, shown for reference.}
    \label{fig:ValidationGPs}
\end{figure}

Parity plots of GP and UT\&C model predictions on the remaining $15\%$ test data, presented in Figure~\ref{fig:ValidationGPs}, indicate reasonable agreement, $R^2$-values above 0.98 for all metrics and with RMSEs that are small relative to the respective metric ranges: $0.119$~$^\circ$C RMSE against a $5.167$~$^\circ$C range in $T_\mathrm{peak}$, $0.034$~$^\circ$C RMSE against a $2.647$~$^\circ$C range in $T_\mathrm{night}$, and $0.167$~$^\circ$C RMSE against a $7.859$~$^\circ$C range in $\mathrm{UTCI}_\mathrm{peak}$. These results demonstrate that the surrogate models are sufficiently accurate to support the subsequent sensitivity analysis and parameter exploration.

With the retrained GPs, the approximated Pareto front was sampled by minimizing the augmented Tchebycheff cost function for varying weight vectors $\boldsymbol{\lambda}$. 10,000 $\boldsymbol{\lambda}$ were sampled and minimized, yielding 6,668 non-dominated objective vectors. Figure~\ref{fig:ParetoScatter} illustrates the true non-dominated and dominated objective vectors evaluated by the UT\&C model in order to generate the training dataset for the GPs, alongside the 6,668 GP-sampled non-dominated vectors, which remain subject to GP uncertainty. These GP-sampled points are colored by their posterior standard deviation to visualize the approximation's confidence.

The approximated Pareto front shows a clear trade-off between $T_{\mathrm{peak}}$ and $T_{\mathrm{night}}$. The relationship between $\text{UTCI}_{\mathrm{peak}}$ and $T_{\mathrm{night}}$ is more varied; although some conflict occurs, there are areas where both objectives can be minimized with minimal impact on the other. However, these areas also exhibit the highest $\sigma$, marking them as the most uncertain regions of the GP approximation. While $\text{UTCI}_{\mathrm{peak}}$ and $T_{\mathrm{peak}}$ generally follow a simultaneous downward trend, a small trade-off appears in the lower temperature ranges, suggesting that both cannot be decreased indefinitely at the same time.

A visual comparison of the GP-based Pareto front approximations obtained for all ten optimization runs is provided in the Appendix (Figure~\ref{fig:10runGPPareto}).

\paragraph{Comparison with a non-adaptive baseline}

To assess the benefit of adaptive sampling, the proposed BO framework was compared against a purely space-filling LHS approach using an identical computational budget. In place of the 50 initial and 94 adaptive samples, 144 design points were generated solely via LHS. The hypervolume of the resulting non-dominated set, computed using the same fixed reference point in every run, was 690.22, which was substantially lower than that achieved by the BO framework (mean: 911.77, min: 884.59 across ten runs). This difference is expected, as the LHS samples are distributed across the design space without explicitly targeting the Pareto-optimal region, whereas the Tchebycheff-based BO iteratively directs evaluations towards promising trade-offs.

This comparison was further extended to the GP-approximated Pareto fronts generated using 10,000 $\boldsymbol{\lambda}$ values. The adaptive BO framework yielded a mean hypervolume of 945.45 (ranging from 903.68 to 968.73), whereas the space-filling baseline resulted in a hypervolume of 742.66. These results indicate that the superior approximation of the Pareto front is not merely a function of the number of model evaluations, but rather a result of the BO adaptive sampling strategy, which more effectively allocates evaluations to promising regions of the design space.

Overall, the optimization identified a robust set of non-dominated objective vectors that reveal a clear trade-off between daytime heat mitigation and nocturnal cooling, providing the basis for the parameter-specific interpretation presented in the following section.

\subsection{Evaluation of heat mitigation strategies}\label{sec:evaluation}
This section examines the parameters driving the observed trade-off between daytime and nocturnal thermal performance to identify which are actionable as adaptation strategies. To characterize parameter influence across the full design space, we employ global sensitivity indices and an interactive surrogate-based exploration, complementing the patterns identified within the non-dominated set.

Both the sensitivity indices and the interactive exploration rely on predictions from three separate GPs (one per thermal metric), described and validated above. 

\begin{figure}[htbp]
    \centering
    \includegraphics[width=1\linewidth]{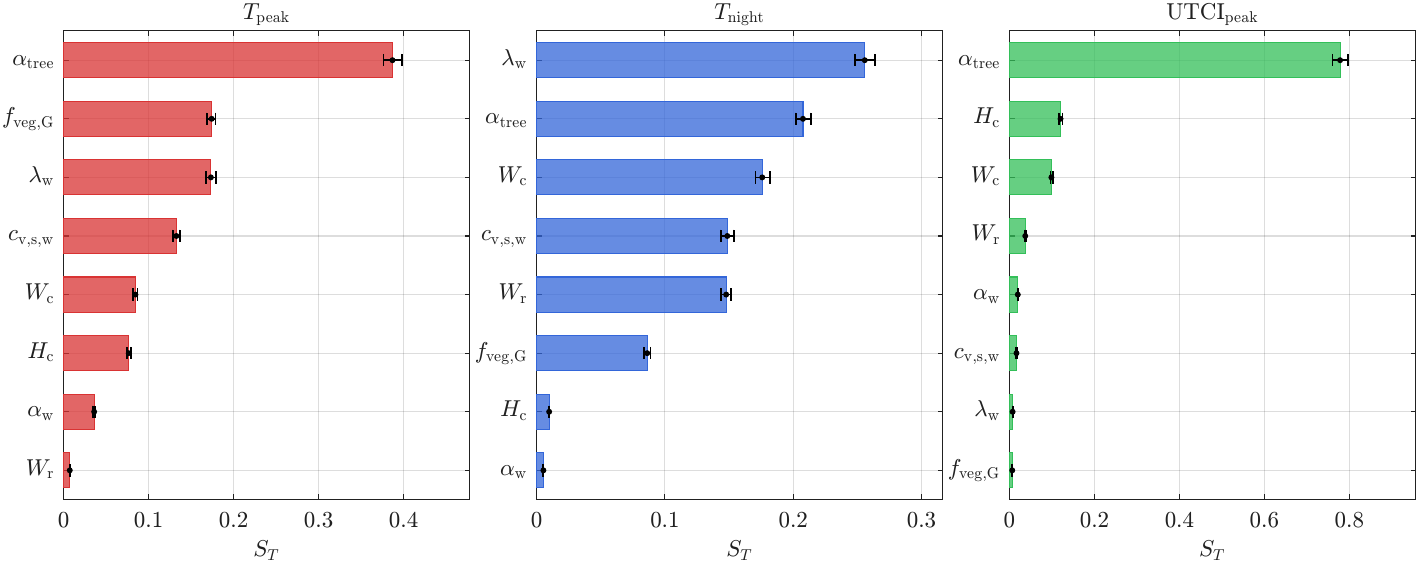}
    \caption{Total-order Sobol sensitivity indices $S_T$ for the three thermal metrics with respect to the eight parameters $H_\mathrm{c}$, $W_\mathrm{c}$, $W_\mathrm{r}$, $f_\mathrm{veg,G}$, $\alpha_\mathrm{tree}$,  $\alpha_\mathrm{w}$, $\lambda_\mathrm{w}$, $c_\mathrm{v,s,w}$, estimated via GPs trained on all of the 144 samples. Indices are computed using the Saltelli estimator \citep{saltelli_variance_2010} (Eq.~\ref{eq:total_sobol}) with N=16,000 quasi-random samples and bootstrapped confidence intervals (200 replicates, error bars, 95\% CI).}
    \label{fig:Sobol}
\end{figure}

\begin{figure}[htbp]
    \centering
    \includegraphics[width=1\linewidth]{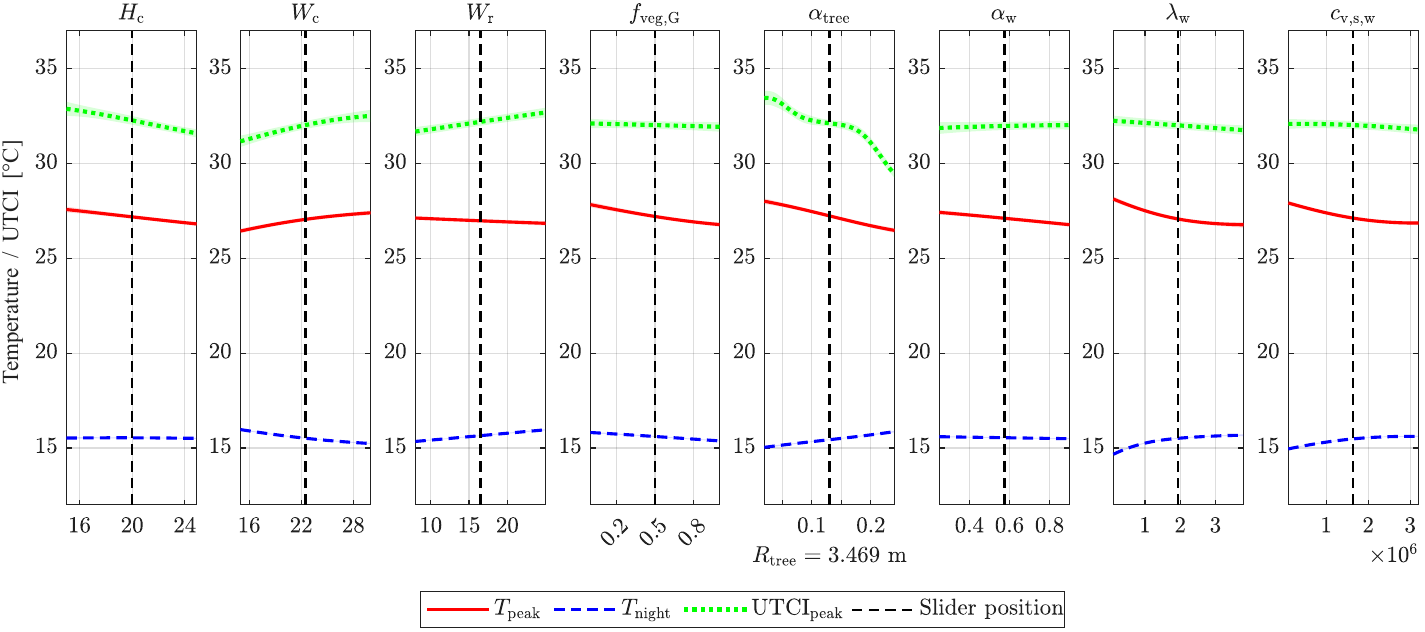}
    \caption{Screenshot of the \emph{Urban Climate Design Explorer} \citep{walter_respace_interaction_2026}. Shaded areas indicate the 95\% prediction intervals, which may be more apparent when interacting with the tool. The vertical black dashed lines indicate the parameter values selected by the sliders. In the interactive version, the slider positions can be moved within their respective parameter ranges to explore parameter interactions and trade-offs between objectives. 
}
    \label{fig:Slider}
\end{figure}

Total-order Sobol indices $S_T$ (Figure~\ref{fig:Sobol}) indicate that tree crown radius scaling $\alpha_\mathrm{tree}$ is among the most influential parameters across all investigated thermal metrics. Wall thermal conductivity $\lambda_\mathrm{w}$ shows a particularly strong influence on $T_\mathrm{night}$, exceeding the influence of $\alpha_\mathrm{tree}$ for $T_\mathrm{night}$. While the vegetated ground fraction $f_\mathrm{\text{veg,G}}$ shows moderate sensitivity for $T_\mathrm{peak}$ and $T_\mathrm{night}$, its impact on $\mathrm{UTCI}_\mathrm{peak}$ is negligible. Canyon geometry shows moderate sensitivity. As the sensitivity analysis only quantifies overall variance contributions without providing directional or functional information, the sign, form and interdependencies of these relationships are further examined using the \emph{Urban Climate Design Explorer} (Fig.~\ref{fig:Slider}) \citep{walter_respace_interaction_2026} 

Fig.~\ref{fig:Slider} makes parameter-response relationships directly visible. Each subplot shows the modeled response of $T_\mathrm{peak}$, $T_\mathrm{night}$, and $\mathrm{UTCI}_\mathrm{peak}$ to variations in one of the eight input parameters, while all other parameters are held constant at their reference values. Larger Sobol indices are reflected in steeper or more strongly non-linear response curves. In the interactive implementation, individual parameters can be varied while all others are held fixed, triggering an immediate update of the response surfaces. This enables exploration of parameter interactions and trade-offs between competing objectives. Based on these parameter–response structures, the following subsections interpret the underlying physical mechanisms and discuss their implications for urban heat mitigation strategies.

\paragraph{Geometric configuration}

A higher canyon ($H_\mathrm{\text{c}}$) generally reduces $T_{\mathrm{peak}}$ and $\mathrm{UTCI}_{\mathrm{peak}}$ by limiting incoming solar radiation. However, this benefit may be partially offset by reduced ventilation within the urban canyon. At night, restricted air exchange can inhibit the removal of stored heat. In addition, taller buildings provide larger wall surface areas that absorb solar radiation during the day and release it after sunset, contributing to the urban heat island effect. While UT\&C accounts for aerodynamic resistance within the canyon, it is questionable whether the model is capable of reproducing such effects realistically considering the small effect of canyon height on $T_{\mathrm{night}}$ in the surrogate model and the simplification made regarding turbulent heat exchange in UT\&C \citep{mateen_large_2025}. 

In contrast, canyon width $W_\mathrm{\text{c}}$ exerts a stronger and ambiguous influence on $T_\mathrm{\text{night}}$. At night, wider canyons promote cooling, which may be attributed to reduced longwave radiation trapping and enhanced ventilation. However, during the day, increased exposure to solar radiation can counteract the cooling benefits. The impact of the roof width $W_\mathrm{\text{r}}$ on $T_{\mathrm{peak}}$ is interlinked with canyon width $W_\mathrm{\text{c}}$. While within all types of geometric configurations, the width of the roof increases $\mathrm{UTCI}_{\mathrm{peak}}$ and slightly increases $T_{\mathrm{night}}$, in a narrow canyon, an increasing roof width slightly reduces $T_{\mathrm{peak}}$. Contrarily, a higher roof width increases $T_{\mathrm{peak}}$ in a wide canyon, an effect, that diminishes with increasing building height $H_\mathrm{\text{c}}$.

\paragraph{Vegetation characteristics}

Both ground vegetation and tree characteristics have a substantial impact on thermal properties. High ground vegetation fractions $f_\mathrm{\text{veg,G}}$ substantially reduce daytime and nighttime air temperature as well as $\mathrm{UTCI}_{\mathrm{peak}}$. This impact arises from increasing evapotranspiration and the decrease in the amount of heat stored within the ground layer compared to impervious surfaces. The influence of ground vegetation decreases with increasing tree crown radius ($\alpha_\mathrm{\text{tree}}$), as enhanced canopy shading reduces the available energy at ground level and shifts the dominant evapotranspiration contribution towards the tree crown layer.

With increasing tree radius $\alpha_\mathrm{\text{tree}}$, the GPs indicate a reduction in $T_{\mathrm{peak}}$ and $\mathrm{UTCI}_{\mathrm{peak}}$, as a result of combined transpirative cooling and shading, while a slight increase in $T_{\mathrm{night}}$ was observed. The effect of trees on air temperature within different climates was analyzed by \cite{meili2021trees} who preserved the urban geometry and separated the impact of trees into the effect caused by shading, evapotranspiration, and aerodynamic roughness alteration. They concluded that evapotranspiration, which encompasses transpiration of the tree itself as well as evaporation of water intercepted by leaves, was the main driver in reducing air temperature. Their study found that the isolated interaction between trees and radiation increased air temperature through greater absorption of shortwave radiation and increased sensible heat emission. In the present study, however, increasing tree radius had a greater positive influence on $\mathrm{UTCI}_{\mathrm{peak}}$ than on $T_{\mathrm{peak}}$. $\mathrm{UTCI}_{\mathrm{peak}}$ integrates other variables such as mean radiant temperature and therefore indicates that the shading effect of trees may still improve thermal comfort independently of increases in air temperature. This emphasizes the importance of an integrative examination of thermal metrics beyond air temperature.

$\mathrm{UTCI}_{\mathrm{peak}}$ shows a particularly strong decline with increasing tree radius in narrow and low canyons, confirming the hypothesis that extensive shading of most of the street surface by trees leads to improved thermal comfort. The effect on daytime air temperature is also higher in lower canyons which may be attributed to the larger fraction of the canyon being occupied by trees in more shallow canyons. The turbulent exchange within UT\&C is governed by a simple resistance scheme which incorporates the influence of tree structure on the aerodynamic roughness within the canopy layer \citep{meili2021trees}. In a more shallow canyon, the tree may have a stronger influence on the turbulent exchange of energy resulting in an alteration of the wind profile.

In high and wide canyons, thermal parameters are least affected by varying tree radius $R_\mathrm{\text{tree}}$ which is in line with previous studies \citep{huang_synergistic_2021}. Trees occupy only a small fraction of the canyon volume which results in reduced shading potential and in a comparatively small interaction with the canyon air. However, tree characteristics exhibit a particularly complex behavior. Within a very wide and low street canyon, $\mathrm{UTCI}_{\mathrm{peak}}$ shows a slight local maximum at about 2 m tree radius and slightly lowers with a decreasing tree radius. With an increasing canyon height, this local maximum shifts towards a higher tree radius, reaching about 3.5 m at maximum canyon height and width. UTCI is simulated at the center of the street, so in wide canyons, a small canopy may not yet provide shade there. Only starting from a certain radius the center becomes shaded by the canopy. Moreover, the maximum may suggest that in this particular configuration, competing processes are acting simultaneously. While an increasing radius enhances the shading effect, a certain tree size may reduce urban roughness and reduce canyon ventilation \citep{meili2021trees, zhao_time-evolving_tree_2023}. 

During the night, a higher tree radius $R_\mathrm{\text{tree}}$ slightly increases $T_{\mathrm{night}}$, which was most pronounced in high and narrow canyons. The increasing tree radius reduces the fraction of open sky which in turn leads to more longwave radiation being trapped inside the canyon during the night \citep{wujeska_2020_nighttimecooling}. This effect is in line with the analysis of the sole radiation effect of trees within urban canyons by \cite{meili2021trees}. High and narrow canyons already feature a smaller sky-view factor which leads to a stronger reabsorption of longwave radiation during the night and this effect is intensified by broad tree canopies. 

While there is already a strong connection between canyon geometry and tree radius during the day and night, in reality, this effect may be further impacted by wind blockage effects of the tree canopy. Wind blockage simulation is not yet incorporated within UT\&C model architecture \citep{meili2021trees}. Moreover, we did not alter the tree height which may have had an influence on wind blockage and aerodynamic roughness effects. Especially, when trees are shorter than average building height, they can decrease aerodynamic roughness of the urban canopy and dampen turbulent exchange processes \citep{zhao_time-evolving_tree_2023}.

\paragraph{Surface and material properties}

Compared to geometric and vegetation parameters, variations in surface and material properties exhibit a smaller influence on the objective vectors. Increasing wall albedo $\alpha_{\mathrm{w}}$ reduces both $T_{\mathrm{peak}}$ and $T_{\mathrm{night}}$, while simultaneously increasing $\mathrm{UTCI}_{\mathrm{peak}}$ due to enhanced reflection of shortwave radiation. The reduction in $T_{\mathrm{peak}}$ is most pronounced in tall canyons and slightly stronger in tall, narrow canyons, which may be attributed to the smaller amount of radiation absorbed by building surfaces. The resulting smaller heat storage decreases $T_{\mathrm{night}}$, enhancing nighttime cooling in urban areas. When tested beforehand, roof albedo along with roof vegetation only presented a marginal effect on thermal parameters with a slightly decreasing $\mathrm{UTCI}_{\mathrm{peak}}$ at higher roof albedo. This may be due to the missing interaction of sensible heat from roof surfaces with the air masses within the canyon in UT\&C \citep{meili_supplement_2020}. Past studies on urban adaptation measures indicated a stronger impact of high albedo materials such as 'cool roofs' and roof vegetation on urban microclimate \citep{elnabawi_super_2023}.

A higher volumetric heat capacity $c_\mathrm{\text{v,s,w}}$ of the walls leads to a lower $T_{\mathrm{peak}}$ and $\mathrm{UTCI}_{\mathrm{peak}}$, but also increases $T_{\mathrm{night}}$ because heat from the outside atmosphere enters the wall mass more efficiently. This highlights the trade-off between increasing building energy efficiency which may increase the daytime outside heat exposure during hot weather \citep{heusinger_modeling_2023}. The impact of volumetric heat capacity on air temperature also depends on the thermal conductivity of the wall. A higher heat conductivity $\lambda_{\mathrm{w}}$ increases $T_{\mathrm{night}}$ especially with high $c_\mathrm{\text{v,s,w}}$ because heat stored within the building during the day is released the outer surface and atmosphere more easily. With increasing thermal conductivity, the warming nighttime impact of the volumetric heat capacity intensifies, while the magnitude of the decrease in $T_{\mathrm{peak}}$ and $\mathrm{UTCI}_{\mathrm{peak}}$ remains similar.

The different ways in which the parameters affect the thermal properties depending on the status of the general urban configuration reflect the complex interplay between the parameters. The multi-objective optimization was used to identify parameter configurations reflecting physical mechanisms that collectively reduce absorption of shortwave radiation, heat storage, and enhance cooling through evapotranspiration. The most effective heat mitigation measures to provide a reduction in $T_{\mathrm{peak}}$, $T_{\mathrm{night}}$ and $\mathrm{UTCI}_{\mathrm{peak}}$  depend on the initial urban geometry. While the optimization in the present study was conducted for a particular canyon orientation, a varying canyon orientation would again affect the efficacy of certain adaptation measures due to an alteration to the radiation dynamics. 

\subsection{Limitations of the optimization}\label{sec:limitations}

Some limitations of the present framework should be acknowledged. The optimization is based on the UT\&C model which inherently features some limitations such as the simplified urban geometry and turbulent transport of momentum and heat. Furthermore, the model is not coupled to mesoscale climatic processes and therefore does not capture larger-scale feedback such as changes in downwelling longwave radiation that may dampen or amplify the impact of changing urban configurations. The results should consequently be interpreted primarily in terms of relative sensitivities and trade-offs between heat mitigation strategies rather than exact temperature magnitudes. The optimization is only conducted for a study area in a temperate climate with sufficient water supply on most days. \cite{meili2021trees} discovered a reduced transpirative cooling of trees in dry and hot climates. Under future climatic conditions or if the optimization were to be conducted for more arid regions, water availability is an important factor when evaluating the impact of tree structure on urban climate.

Some expected urban climate dynamics, particularly the nocturnal release of heat stored within the urban fabric and the resulting urban heat island effect, appear to be only weakly represented in the simulations. Noticeable deviations between measured and simulated temperatures occur during the early night hours, when the urban heat island effect is typically most pronounced \citep{OKE1973769}. Experiments with the model configuration, for example by increasing parameters related to internal building mass and heat storage, resulted only in minor changes in nighttime temperatures. Improving the representation of urban heat storage and release processes would therefore be an important direction for future model development.

Additionally, the optimization focuses on three indicators of heat stress, namely the daytime peak temperature $T_{\mathrm{peak}}$, the nighttime minimum temperature $T_{\mathrm{night}}$, and the peak thermal comfort metric $\mathrm{UTCI}_{\mathrm{peak}}$. While these metrics capture important aspects of urban heat exposure, they do not fully represent all dimensions of thermal resilience, such as the duration of heat exposure, seasonal variability, or spatial heterogeneity of pedestrian experience. Economic considerations are not explicitly incorporated in the optimization. Many mitigation measures represented in the parameter space are associated with different implementation and maintenance costs. While the optimization identifies configurations with favorable thermal performance, their practical feasibility ultimately depends on economic constraints, regulatory frameworks, and implementation conditions faced by urban decision makers. 

\section{Summary and Conclusions}

The presented framework demonstrates how surrogate-based optimization can support the systematic exploration of entire parameter spaces regarding climate heat mitigation strategies in urban environments. By linking urban morphology parameters of a street canyon in the mid-sized German city of Braunschweig with multiple indicators of heat stress, the approach reveals the structure of trade-offs between daytime thermal comfort, nocturnal cooling, and radiative heat exposure. It was shown that improvements in daytime thermal conditions are often associated with higher nighttime temperatures.

The substantial temperature ranges spanned across the evaluated configurations demonstrate that urban design decisions can strongly influence thermal conditions during extreme heat events. Under identical forcing conditions, different urban configurations, varying in geometry, vegetation, and surface characteristics, are shown to alter peak canyon air temperature by up to 5.2 $^\circ$C during the day and 2.6 $^\circ$C at night, while UTCI varies by up to 7.9 $^\circ$C. This demonstrates that favorable street morphology can substantially mitigate microclimatic heat stress. From an urban planning perspective, the optimization results indicate that effective heat mitigation strategies feature an increase in roadside vegetation and street trees as well as high-albedo surfaces.

As cities increasingly face more frequent and intense heat events, the framework provides a methodological basis for exploring trade-offs in urban climate adaption strategies under realistic planning constraints. While the present analysis considered the warmest days in August 2021, the framework can be naturally extended to future scenario analysis. This can be achieved, for instance, by combining a microscale climate model with mesoscale forecasting models such as the Weather Research and Forecasting (WRF) model.
Such coupling would provide the possibility to apply future climatic conditions to the optimization approach, with more frequent and intense droughts and heat waves in the future likely affecting the efficacy of adaptation strategies.
The framework of optimization could also be applied to other urban climate models with more detailed representation of urban structure to enable a more quantitative and a spatially explicit evaluation of heat mitigation strategies. Future work could integrate additional decision criteria, including economic cost functions or lifecycle considerations. Such extensions would enable the identification of mitigation strategies that are not only thermally effective but also economically viable for municipalities and planners. Approaches of combined BO, surrogate modeling, and interactive exploration interfaces may help bridge the gap between computationally intensive urban climate simulations and practical decision-making processes in urban planning. 

\section*{Data}
The code used in this study was developed in MATLAB R2025b and will be made publicly available upon publication \citep{MyCode2026}.

\section*{CRediT authorship contribution statement}
Rebekka Walter: Conceptualization, Methodology, Software, Writing - Original Draft, Writing - Review \& Editing, Visualization.
Johanna Gelhaus: Conceptualization, Validation, Writing - Original Draft, Writing - Review \& Editing.
David Anton: Methodology, Writing - Review \& Editing.
Henning Wessels: Conceptualization, Methodology, Writing - Review \& Editing, Supervision, Funding acquisition.
Stephan Weber: Conceptualization, Resources, Writing - Review \& Editing, Supervision.

\section*{Declaration of competing interest}
The authors declare that they have no known competing financial interests or personal relationships that could have appeared to influence the work reported in this paper.

\section*{Funding}
The research project "ReSpace!" is funded by zukunft.niedersachsen, the joint science funding program of the Lower Saxony Ministry of Science and Culture and the Volkswagen Foundation under the TU Braunschweig initiative "Ecoversity" with the funding number 11-76251-2714/2024 (ZN4545).

\section*{Declaration of generative AI and AI-assisted technologies in the manuscript preparation process}
During the preparation of this work, the authors used Claude and ChatGPT to improve wording, and grammar of the manuscript, and to assist with code debugging and optimization. The authors reviewed and critically evaluated all AI-assisted outputs and take full responsibility for the content of the published article.

\newpage

\appendix
\section{Details on UT\&C model verification}\label{sec:verification}

\paragraph{Evaluation of UT\&C predicted air temperature}

\begin{figure}[htbp]
    \centering
    \includegraphics[height=0.45\textheight]{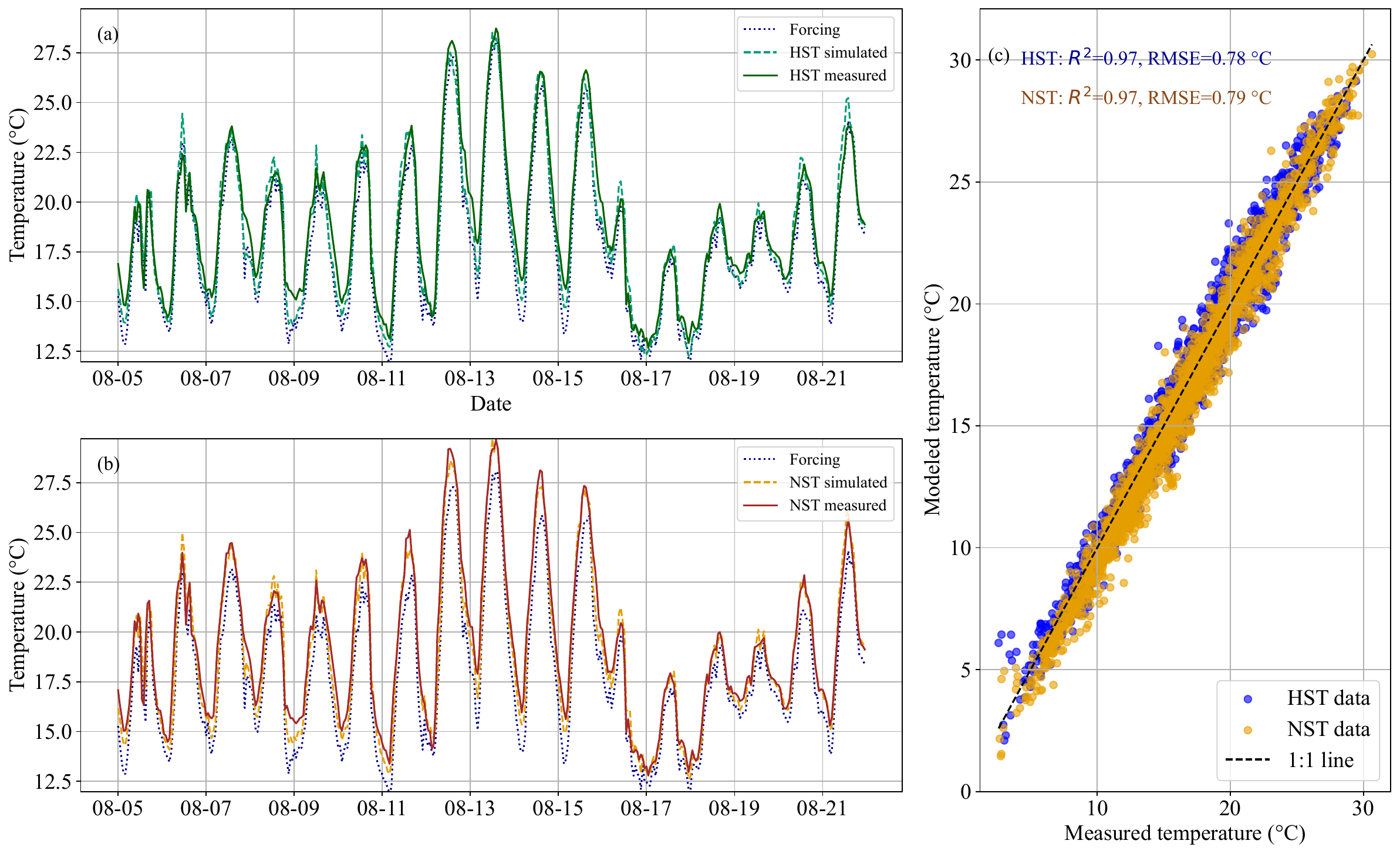}
    \caption{(a) Time series of modeled and measured temperature in mid-August, when the daytime forcing temperature in the study period was highest and (b) comparison of measured and simulated data for the entire study period.}
    \label{fig:temp_validation_graph}
\end{figure}
\begin{figure}[htbp]
    \centering
    \includegraphics[height=0.45\textheight]{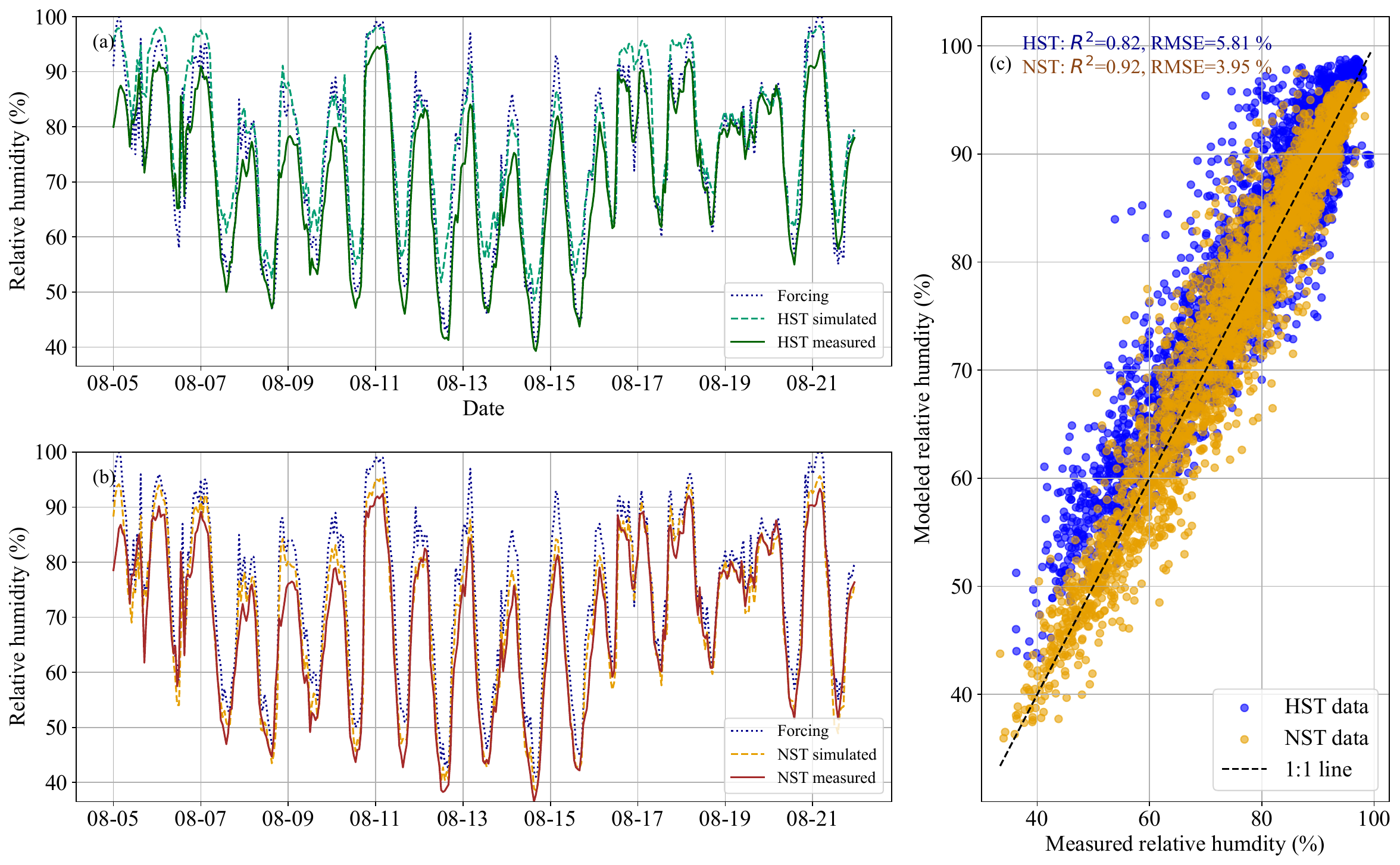}
    \caption{(a) Time series of modeled and measured relative humidity in mid-August, when the daytime forcing temperature in the study period was highest and (b) comparison of measured and simulated data for the entire study period.}
    \label{fig:RH_validation_graph}
\end{figure}

\begin{figure}[htbp]
    \centering
    \includegraphics[width=0.8\linewidth]{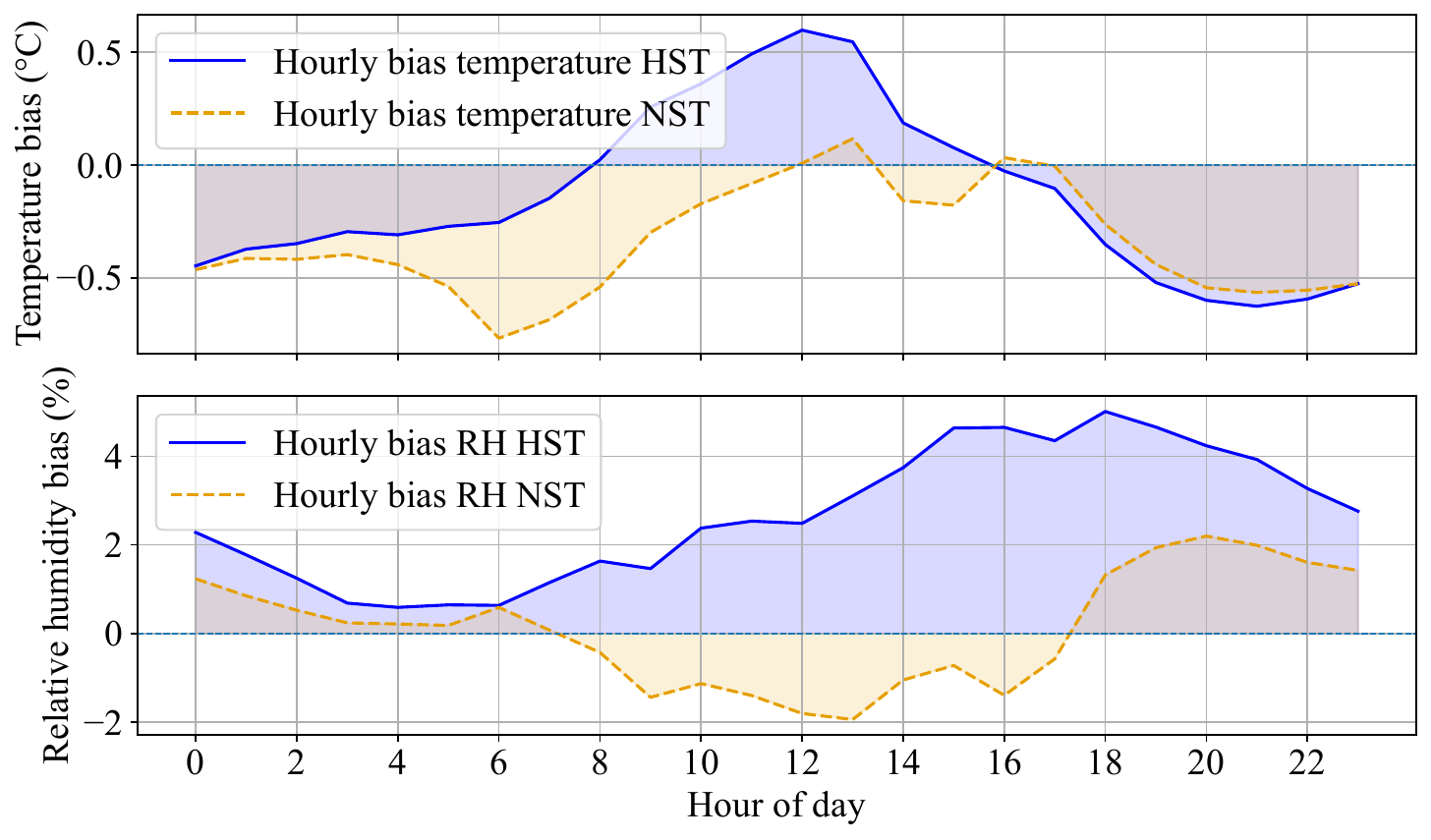}
    \caption{Mean diurnal difference between simulated and measured temperature (top) and relative humidity (bottom) for HST and NST, respectively.}
    \label{fig:hourly_difference}
\end{figure}

The scatter plot of observed and modeled temperature data for the entire study period indicates a good fit between the observed and simulated data, with the UT\&C model capable of explaining the vast majority of the variance in the observed data for both streets (Figure~\ref{fig:temp_validation_graph}c). The mean error indicates a general slight underestimation by the model for both HST and NST while the RMSE indicates a reasonably good overall model performance (Table~\ref{tab:utc_eval_metrics}). During the day, the mean error of simulated air temperatures at HST showed very good agreement with the measurements, while at night, the model tended to underestimate temperatures on average. Considering the mean diurnal variation, the maximum average overestimation by 0.60 $^\circ$C occurred at 12:00, while the highest negative mean error of -0.63 $^\circ$C was at 21:00 (Figure~\ref{fig:hourly_difference}). The simulated air temperatures at NST indicated a slightly stronger underestimation of the measured values during the day and the night. With regard to the mean diurnal variation, a slight overestimation by the model only occured at 12:00 and 13:00. Maximum underestimation by -0.56 $^\circ$C was at 21:00. 

 Compared to the model validation conducted for measurements in Singapore by \cite{meili2020utc}, the model performs better for the two streets in Braunschweig during the day and night. In the study for Singapore, the mean error was at 0.9 $^\circ$C during the day and -1.2 $^\circ$C at night. \cite{chen_effects_2023} analyzed the effect of street trees within different urban configurations with an experimental model setup and discovered that the model performed better regarding temperature within their setup with trees. In this study, the opposite was true, with NST simulations performing slightly better. However, sensor placement may play a crucial role in this study. The slight underestimation of NST temperature during the day may arise from the sensor location within NST on the northern side of the street. The sensor was thereby located more closely to the wall that is exposed to sunlight, while in HST, the sensor was located more closely to the southern street side and may therefore have been shaded by the southern buildings. Moreover, the difference in height between the simulation location and the sensor location may have created a systematic offset for both streets. Another aspect not represented in the model architecture are the gable roofs featured in HST and NST, which interact differently with radiation than flat roofs. Additionally, assumption of albedo, emissivity, and thermal characteristics of building materials may not reflect the real conditions which may account for spatially averaged differences of about 1 $^\circ$C as well \citep{heusinger_modeling_2023}.

 On most days such as August 12 and 13, the simulated peak NST temperature is higher than HST temperature, which reflects the difference in the measurements (Figures~\ref{fig:temp_validation_graph}a and b). Although NST is characterized by little vegetation and thereby expected to heat up more, the measured temperature differences between HST and NST only amount to about 2.5 $^\circ$C even on hot days. The street canyon of NST is much narrower leading to reduced solar radiation entering and heating surfaces. On other days such as August 11, the simulated temperatures for HST and NST are almost equal whereas the observed NST temperature is much higher than HST. The observed variability in model agreement between HST and NST on different days may arise from slight changes in wind direction and amount of incoming shortwave radiation on those days due to the different canyon orientation and sensor placement. The large park east of the street canyons was shown to act as a cold air reservoir and may have had a varying effect depending on wind direction \citep{grunwald_mapping_2019}. 

\paragraph{Evaluation of UT\&C predicted relative humidity}

The temporal courses of measured and modeled RH generally show similar patterns (Figure~\ref{fig:RH_validation_graph}a and b). Despite distinct phases during which much higher nighttime humidity was measured at the rural station than within the street canyon, the temporal course still shows good agreement between modeled and measured data. For example, during the night of August 24-25, the forcing humidity reached values of up to 100\%, whereas both measured and simulated RH within the street canyon remained similar, ranging between 85 and 90\%. During other nights, such as August 27-28, RH measured at the rural site and within the street canyon was similarly high, and the simulation likewise showed good agreement with the observations. The comparison of measured and simulated RH shows stronger scattering than the temperature values, indicating a slightly inferior model performance for RH compared to air temperature (Figure~\ref{fig:RH_validation_graph}c). However, the majority of the variance in the observed data can still be explained by the model. The mean error and RMSE indicate moderate deviations between modeled and measured RH that are more pronounced for HST.

During the night, the modeled RH slightly overestimates the measurements for both streets with typical deviations by around 3-5 \%, but performed slightly better than at daytime. During the day, the model also slightly overestimated the measured RH values in HST whereas for NST, the model underestimated daytime RH measurements. The negative error may be due to the fact that no vegetation is implemented in the model setup of NST. While this reflects the setup of the street canyon for the model design, the processes within the real street canyon are not uncoupled from vegetated gardens and streets nearby. The transpiration performed by the vegetation nearby can increase the humidity within the street canyon through the exchange of air masses. 

The strongest mean underestimation for NST occurred at 13:00 with -1.94 \% (Figure~\ref{fig:hourly_difference}). The maximum overestimation of the measured values occurred in the early evening hours between 18:00 and 20:00, when incoming shortwave radiation decreased rapidly, and was more pronounced for HST. It is possible that the closing of stomata, which occurs with decreasing shortwave radiation, is simulated too slowly resulting in a lag in the decreasing transpiration between simulated and measured data. The comparison conducted by \cite{chen_effects_2023} also indicated a daytime underestimation in the setup without trees while RH was overestimated with trees present. Along with other studies, they attributed the overestimation of RH species-specific physiological reactions to temperature and water stress \citep{tan_transpiration_2020} and to an underestimation of the turbulent exchange of heat and humidity \citep{grimmond_international_2010, huang_synergistic_2021}.

\section{Details on reference run}\label{sec:144timeseries}

\begin{figure}[H]
    \centering
    \includegraphics[width=1\linewidth]{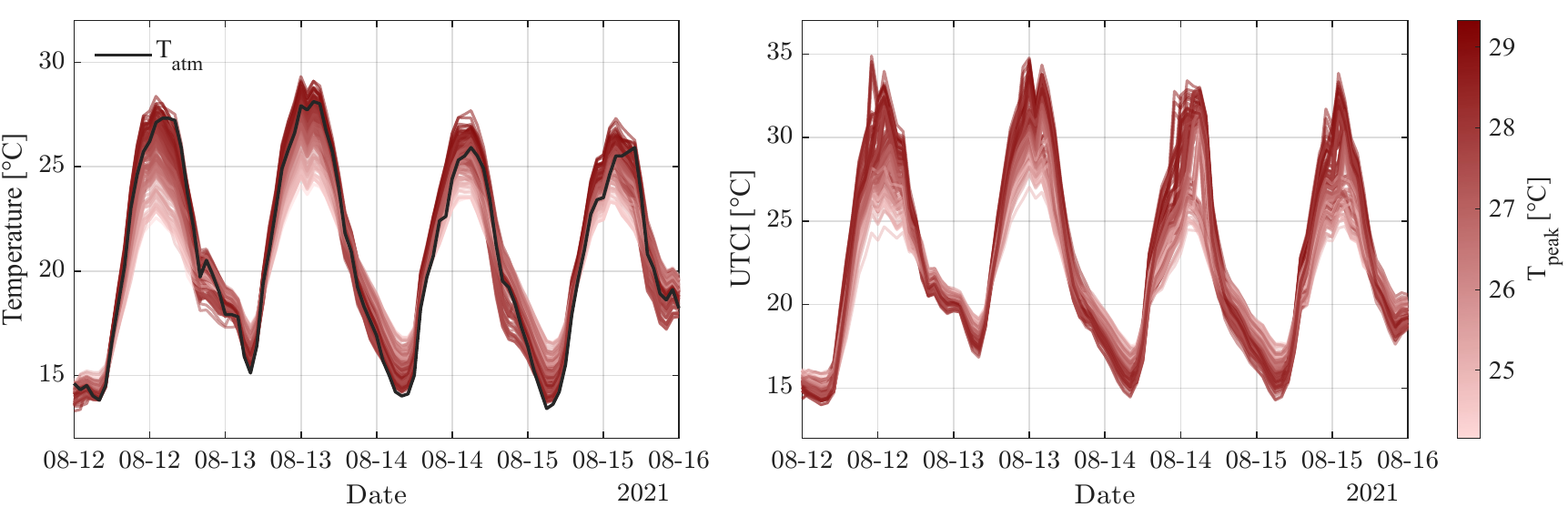}
    \caption{Temperature and UTCI timeseries for all 144 evaluated UT\&C runs from BO run 8, colored by $T_\mathrm{peak}$}
    \label{fig:144TempUTCItimeseries}
\end{figure}

\section{Details on 10 run convergence} \label{sec:DensePareto}

\begin{table}[htbp]
\centering
\caption{Summary of the ten independent Bayesian optimization runs.}
\label{tab:multi_runs}
\begin{tabular}{cccc}
\hline
Run & BO iterations & Final hypervolume &  Non-dominated objective vectors \\
\hline
1  & 118 & 921.39 & 33 \\
2  & 97  & 918.99 & 25 \\
3  & 124 & 900.75 & 38 \\
4  & 105 & 920.78 & 33 \\
5  & 91  & 917.46 & 28 \\
6  & 85  & 907.29 & 36 \\
7  & 98  & 884.59 & 30 \\
8  & 94  & 914.97 & 33 \\
9  & 83  & 907.93 & 28 \\
10 & 87  & 923.52 & 32 \\
\hline
\end{tabular}
\end{table}

\begin{table}[htbp]
\centering
\caption{Statistical summary of the ten Bayesian optimization runs.}
\label{tab:summultiruns}
\begin{tabular}{lccccc}
\hline
Metric & Mean & Std. dev. & CV (\%) & Minimum & Maximum \\
\hline
BO iterations & 98.2 & 13.77 & 14.02 & 83 & 124 \\
Final hypervolume & 911.77 & 12.02 & 1.32 & 884.59 & 923.52 \\
Non-dominated objective vectors & 31.6 & 3.92 & 12.41 & 25 & 38 \\
\hline
\end{tabular}
\end{table}

\begin{figure}[htbp]
    \centering
    \includegraphics[width=1\linewidth]{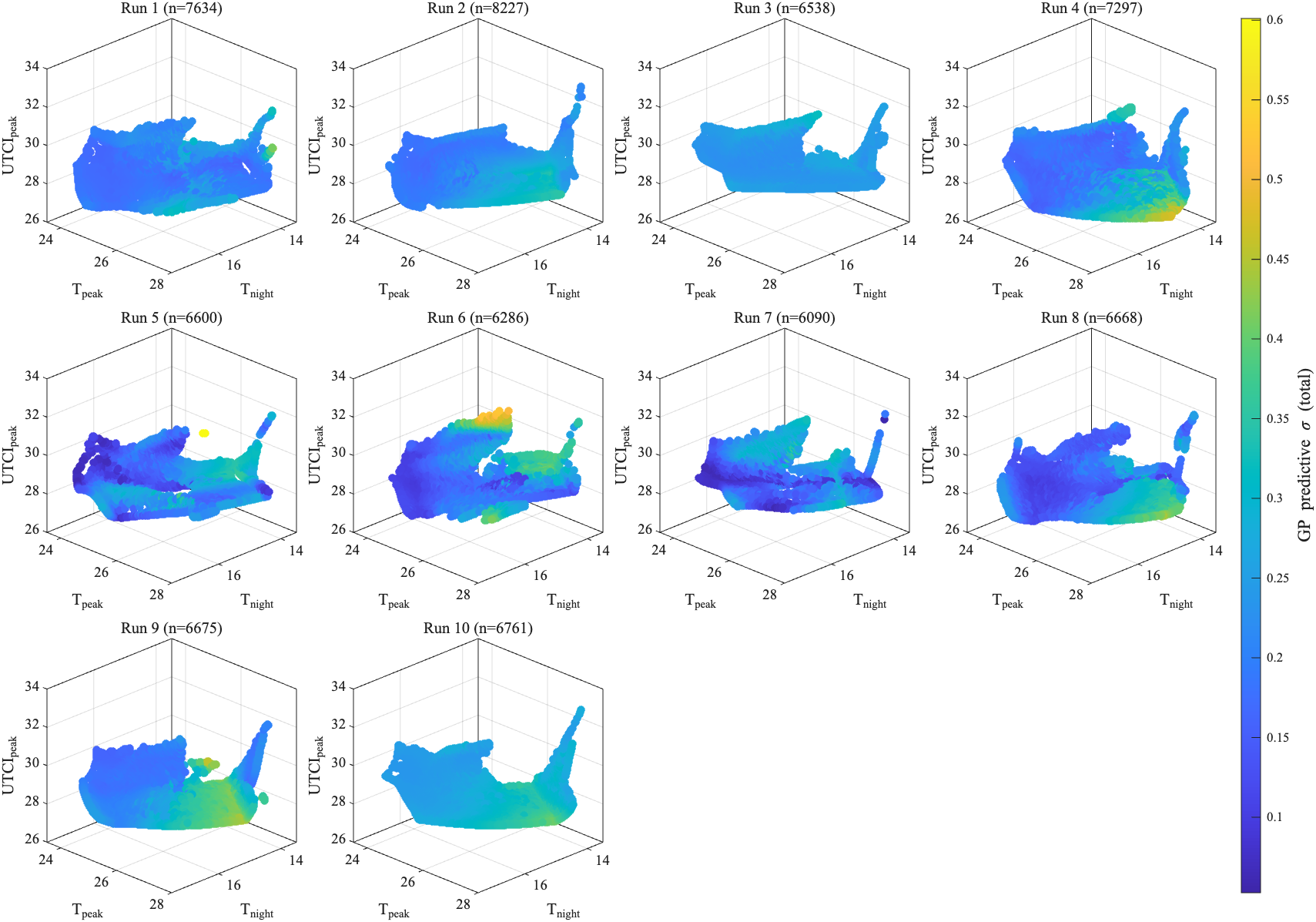}
    \caption{Comparison of the GP-based Pareto front approximations obtained from ten independent BO runs. In each BO run, a distinct sequence of weighting vectors $\boldsymbol{\lambda}$ was generated using a different random seed, resulting in different scalarization directions and, consequently, different sets of evaluated design points and fitted GPs. After convergence of each BO procedure, the augmented Tchebycheff scalarization was applied to 10,000 common weighting vectors $\boldsymbol{\lambda}$ using only the three corresponding GPs (one per thermal metric), without further evaluation of the UT\&C model. Thus, differences between the approximated Pareto fronts originate from the different GP models obtained in the independent BO runs rather than from the weighting vectors used for this final approximation. The number $n$ denotes the number of unique objective vectors obtained from the 10,000 weighting vectors $\boldsymbol{\lambda}$; for example, $n=7,634$ in run~1 indicates that 7,634 unique non-dominated objective vectors were identified for 10,000 $\boldsymbol{\lambda}$. The markers denote the resulting GP-based non-dominated objective vectors and are colored according to the combined posterior predictive standard deviation of the three GP models, calculated as the Euclidean norm of their individual predictive standard deviations. Dark blue indicates higher predictive certainty. (See also: Post-processing and visualization, \ref{sec:pareto}}
    \label{fig:10runGPPareto}
\end{figure}


\clearpage

\begin{thebibliography}{49}
\expandafter\ifx\csname natexlab\endcsname\relax\def\natexlab#1{#1}\fi
\providecommand{\url}[1]{\texttt{#1}}
\providecommand{\href}[2]{#2}
\providecommand{\path}[1]{#1}
\providecommand{\DOIprefix}{doi:}
\providecommand{\ArXivprefix}{arXiv:}
\providecommand{\URLprefix}{URL: }
\providecommand{\Pubmedprefix}{pmid:}
\providecommand{\doi}[1]{\href{http://dx.doi.org/#1}{\path{#1}}}
\providecommand{\Pubmed}[1]{\href{pmid:#1}{\path{#1}}}
\providecommand{\bibinfo}[2]{#2}
\ifx\xfnm\relax \def\xfnm[#1]{\unskip,\space#1}\fi
\bibitem[{Bröde et~al.(2012)Bröde, Fiala, Błażejczyk, Holmér, Jendritzky, Kampmann, Tinz and Havenith}]{brodeDerivingOperationalProcedure2012}
\bibinfo{author}{Bröde, P.}, \bibinfo{author}{Fiala, D.}, \bibinfo{author}{Błażejczyk, K.}, \bibinfo{author}{Holmér, I.}, \bibinfo{author}{Jendritzky, G.}, \bibinfo{author}{Kampmann, B.}, \bibinfo{author}{Tinz, B.}, \bibinfo{author}{Havenith, G.}, \bibinfo{year}{2012}.
\newblock \bibinfo{title}{Deriving the operational procedure for the universal thermal climate index ({UTCI})}.
\newblock \bibinfo{journal}{International Journal of Biometeorology} \bibinfo{volume}{56}, \bibinfo{pages}{481--494}.
\newblock \URLprefix \url{https://doi.org/10.1007/s00484-011-0454-1}, \DOIprefix\doi{10.1007/s00484-011-0454-1}.
\bibitem[{Calvin et~al.(2023)Calvin, Dasgupta, Krinner, Mukherji, Thorne, Trisos, Romero, Aldunce, Barrett, Blanco, Cheung, Connors, Denton, Diongue-Niang, Dodman, Garschagen, Geden, Hayward, Jones, Jotzo, Krug, Lasco, Lee, Masson-Delmotte, Meinshausen, Mintenbeck, Mokssit, Otto, Pathak, Pirani, Poloczanska, Pörtner, Revi, Roberts, Roy, Ruane, Skea, Shukla, Slade, Slangen, Sokona, Sörensson, Tignor, Van~Vuuren, Wei, Winkler, Zhai, Zommers, Hourcade, Johnson, Pachauri, Simpson, Singh, Thomas, Totin, Alegría, Armour, Bednar-Friedl, Blok, Cissé, Dentener, Eriksen, Fischer, Garner, Guivarch, Haasnoot, Hansen, Hauser, Hawkins, Hermans, Kopp, Leprince-Ringuet, Lewis, Ley, Ludden, Niamir, Nicholls, Some, Szopa, Trewin, Van Der~Wijst, Winter, Witting, Birt and Ha}]{lee_ipcc_2023}
\bibinfo{author}{Calvin, K.}, \bibinfo{author}{Dasgupta, D.}, \bibinfo{author}{Krinner, G.}, \bibinfo{author}{Mukherji, A.}, \bibinfo{author}{Thorne, P.W.}, \bibinfo{author}{Trisos, C.}, \bibinfo{author}{Romero, J.}, \bibinfo{author}{Aldunce, P.}, \bibinfo{author}{Barrett, K.}, \bibinfo{author}{Blanco, G.}, \bibinfo{author}{Cheung, W.W.}, \bibinfo{author}{Connors, S.}, \bibinfo{author}{Denton, F.}, \bibinfo{author}{Diongue-Niang, A.}, \bibinfo{author}{Dodman, D.}, \bibinfo{author}{Garschagen, M.}, \bibinfo{author}{Geden, O.}, \bibinfo{author}{Hayward, B.}, \bibinfo{author}{Jones, C.}, \bibinfo{author}{Jotzo, F.}, \bibinfo{author}{Krug, T.}, \bibinfo{author}{Lasco, R.}, \bibinfo{author}{Lee, Y.Y.}, \bibinfo{author}{Masson-Delmotte, V.}, \bibinfo{author}{Meinshausen, M.}, \bibinfo{author}{Mintenbeck, K.}, \bibinfo{author}{Mokssit, A.}, \bibinfo{author}{Otto, F.E.}, \bibinfo{author}{Pathak, M.}, \bibinfo{author}{Pirani, A.}, \bibinfo{author}{Poloczanska, E.}, \bibinfo{author}{Pörtner, H.O.},
  \bibinfo{author}{Revi, A.}, \bibinfo{author}{Roberts, D.C.}, \bibinfo{author}{Roy, J.}, \bibinfo{author}{Ruane, A.C.}, \bibinfo{author}{Skea, J.}, \bibinfo{author}{Shukla, P.R.}, \bibinfo{author}{Slade, R.}, \bibinfo{author}{Slangen, A.}, \bibinfo{author}{Sokona, Y.}, \bibinfo{author}{Sörensson, A.A.}, \bibinfo{author}{Tignor, M.}, \bibinfo{author}{Van~Vuuren, D.}, \bibinfo{author}{Wei, Y.M.}, \bibinfo{author}{Winkler, H.}, \bibinfo{author}{Zhai, P.}, \bibinfo{author}{Zommers, Z.}, \bibinfo{author}{Hourcade, J.C.}, \bibinfo{author}{Johnson, F.X.}, \bibinfo{author}{Pachauri, S.}, \bibinfo{author}{Simpson, N.P.}, \bibinfo{author}{Singh, C.}, \bibinfo{author}{Thomas, A.}, \bibinfo{author}{Totin, E.}, \bibinfo{author}{Alegría, A.}, \bibinfo{author}{Armour, K.}, \bibinfo{author}{Bednar-Friedl, B.}, \bibinfo{author}{Blok, K.}, \bibinfo{author}{Cissé, G.}, \bibinfo{author}{Dentener, F.}, \bibinfo{author}{Eriksen, S.}, \bibinfo{author}{Fischer, E.}, \bibinfo{author}{Garner, G.}, \bibinfo{author}{Guivarch, C.},
  \bibinfo{author}{Haasnoot, M.}, \bibinfo{author}{Hansen, G.}, \bibinfo{author}{Hauser, M.}, \bibinfo{author}{Hawkins, E.}, \bibinfo{author}{Hermans, T.}, \bibinfo{author}{Kopp, R.}, \bibinfo{author}{Leprince-Ringuet, N.}, \bibinfo{author}{Lewis, J.}, \bibinfo{author}{Ley, D.}, \bibinfo{author}{Ludden, C.}, \bibinfo{author}{Niamir, L.}, \bibinfo{author}{Nicholls, Z.}, \bibinfo{author}{Some, S.}, \bibinfo{author}{Szopa, S.}, \bibinfo{author}{Trewin, B.}, \bibinfo{author}{Van Der~Wijst, K.I.}, \bibinfo{author}{Winter, G.}, \bibinfo{author}{Witting, M.}, \bibinfo{author}{Birt, A.}, \bibinfo{author}{Ha, M.}, \bibinfo{year}{2023}.
\newblock \bibinfo{title}{{IPCC}, 2023: Climate Change 2023: Synthesis Report. Contribution of Working Groups I, {II} and {III} to the Sixth Assessment Report of the Intergovernmental Panel on Climate Change [Core Writing Team, H. Lee and J. Romero (eds.)]. {IPCC}, Geneva, Switzerland.}
\newblock \bibinfo{type}{Technical Report}. Intergovernmental Panel on Climate Change ({IPCC}).
\newblock \URLprefix \url{https://www.ipcc.ch/report/ar6/syr/}, \DOIprefix\doi{10.59327/IPCC/AR6-9789291691647}. \bibinfo{note}{edition: First}.
\bibitem[{Chen et~al.(2023)Chen, Meili, Fatichi, Hang, Tan and Yuan}]{chen_effects_2023}
\bibinfo{author}{Chen, T.}, \bibinfo{author}{Meili, N.}, \bibinfo{author}{Fatichi, S.}, \bibinfo{author}{Hang, J.}, \bibinfo{author}{Tan, P.Y.}, \bibinfo{author}{Yuan, C.}, \bibinfo{year}{2023}.
\newblock \bibinfo{title}{Effects of tree plantings with varying street aspect ratios on the thermal environment using a mechanistic urban canopy model}.
\newblock \bibinfo{journal}{Building and Environment} \bibinfo{volume}{246}, \bibinfo{pages}{111006}.
\newblock \URLprefix \url{https://linkinghub.elsevier.com/retrieve/pii/S0360132323010338}, \DOIprefix\doi{10.1016/j.buildenv.2023.111006}.
\bibitem[{Collette and Siarry(2004)}]{collette_multiobjective_2004}
\bibinfo{author}{Collette, Y.}, \bibinfo{author}{Siarry, P.}, \bibinfo{year}{2004}.
\newblock \bibinfo{title}{Multiobjective Optimization: Principles and Case Studies}.
\newblock Decision Engineering, \bibinfo{publisher}{Springer Berlin Heidelberg}.
\newblock \URLprefix \url{https://link.springer.com/10.1007/978-3-662-08883-8}, \DOIprefix\doi{10.1007/978-3-662-08883-8}.
\bibitem[{{Deutscher Wetterdienst}(2026)}]{dwd_cdc_climate_data}
\bibinfo{author}{{Deutscher Wetterdienst}}, \bibinfo{year}{2026}.
\newblock \bibinfo{title}{Climate data center (cdc): Observations germany climate}.
\newblock \URLprefix \url{https://opendata.dwd.de/climate_environment/CDC/observations_germany/climate/}. \bibinfo{note}{accessed: 2026-05-29}.
\bibitem[{Elnabawi et~al.(2023)Elnabawi, Hamza and Raveendran}]{elnabawi_super_2023}
\bibinfo{author}{Elnabawi, M.H.}, \bibinfo{author}{Hamza, N.}, \bibinfo{author}{Raveendran, R.}, \bibinfo{year}{2023}.
\newblock \bibinfo{title}{‘super cool roofs’: Mitigating the {UHI} effect and enhancing urban thermal comfort with high albedo-coated roofs}.
\newblock \bibinfo{journal}{Results in Engineering} \bibinfo{volume}{19}, \bibinfo{pages}{101269}.
\newblock \URLprefix \url{https://linkinghub.elsevier.com/retrieve/pii/S2590123023003961}, \DOIprefix\doi{10.1016/j.rineng.2023.101269}.
\bibitem[{Fatichi et~al.(2012)Fatichi, Ivanov and Caporali}]{fatichi_mechanistic_2012}
\bibinfo{author}{Fatichi, S.}, \bibinfo{author}{Ivanov, V.Y.}, \bibinfo{author}{Caporali, E.}, \bibinfo{year}{2012}.
\newblock \bibinfo{title}{A mechanistic ecohydrological model to investigate complex interactions in cold and warm water‐controlled environments: 1. theoretical framework and plot‐scale analysis}.
\newblock \bibinfo{journal}{Journal of Advances in Modeling Earth Systems} \bibinfo{volume}{4}, \bibinfo{pages}{2011MS000086}.
\newblock \URLprefix \url{https://agupubs.onlinelibrary.wiley.com/doi/10.1029/2011MS000086}, \DOIprefix\doi{10.1029/2011MS000086}.
\bibitem[{Faymonville et~al.(2026)Faymonville, Heusinger and Weber}]{FAYMONVILLE2026102956}
\bibinfo{author}{Faymonville, L.}, \bibinfo{author}{Heusinger, J.}, \bibinfo{author}{Weber, S.}, \bibinfo{year}{2026}.
\newblock \bibinfo{title}{Combining shortwave reflectivity and evaporation in a novel rooftop solution for urban cooling}.
\newblock \bibinfo{journal}{Urban Climate} \bibinfo{volume}{67}, \bibinfo{pages}{102956}.
\newblock \URLprefix \url{https://www.sciencedirect.com/science/article/pii/S2212095526001872}, \DOIprefix\doi{https://doi.org/10.1016/j.uclim.2026.102956}.
\bibitem[{Grimmond et~al.(2010)Grimmond, Blackett, Best, Barlow, Baik, Belcher, Bohnenstengel, Calmet, Chen, Dandou, Fortuniak, Gouvea, Hamdi, Hendry, Kawai, Kawamoto, Kondo, Krayenhoff, Lee, Loridan, Martilli, Masson, Miao, Oleson, Pigeon, Porson, Ryu, Salamanca, Shashua-Bar, Steeneveld, Tombrou, Voogt, Young and Zhang}]{grimmond_international_2010}
\bibinfo{author}{Grimmond, C.S.B.}, \bibinfo{author}{Blackett, M.}, \bibinfo{author}{Best, M.J.}, \bibinfo{author}{Barlow, J.}, \bibinfo{author}{Baik, J.J.}, \bibinfo{author}{Belcher, S.E.}, \bibinfo{author}{Bohnenstengel, S.I.}, \bibinfo{author}{Calmet, I.}, \bibinfo{author}{Chen, F.}, \bibinfo{author}{Dandou, A.}, \bibinfo{author}{Fortuniak, K.}, \bibinfo{author}{Gouvea, M.L.}, \bibinfo{author}{Hamdi, R.}, \bibinfo{author}{Hendry, M.}, \bibinfo{author}{Kawai, T.}, \bibinfo{author}{Kawamoto, Y.}, \bibinfo{author}{Kondo, H.}, \bibinfo{author}{Krayenhoff, E.S.}, \bibinfo{author}{Lee, S.H.}, \bibinfo{author}{Loridan, T.}, \bibinfo{author}{Martilli, A.}, \bibinfo{author}{Masson, V.}, \bibinfo{author}{Miao, S.}, \bibinfo{author}{Oleson, K.}, \bibinfo{author}{Pigeon, G.}, \bibinfo{author}{Porson, A.}, \bibinfo{author}{Ryu, Y.H.}, \bibinfo{author}{Salamanca, F.}, \bibinfo{author}{Shashua-Bar, L.}, \bibinfo{author}{Steeneveld, G.J.}, \bibinfo{author}{Tombrou, M.}, \bibinfo{author}{Voogt, J.}, \bibinfo{author}{Young,
  D.}, \bibinfo{author}{Zhang, N.}, \bibinfo{year}{2010}.
\newblock \bibinfo{title}{The international urban energy balance models comparison project: First results from phase 1}.
\newblock \bibinfo{journal}{Journal of Applied Meteorology and Climatology} \bibinfo{volume}{49}, \bibinfo{pages}{1268--1292}.
\newblock \URLprefix \url{http://journals.ametsoc.org/doi/10.1175/2010JAMC2354.1}, \DOIprefix\doi{10.1175/2010JAMC2354.1}.
\bibitem[{Grunwald et~al.(2019)Grunwald, Kossmann and Weber}]{grunwald_mapping_2019}
\bibinfo{author}{Grunwald, L.}, \bibinfo{author}{Kossmann, M.}, \bibinfo{author}{Weber, S.}, \bibinfo{year}{2019}.
\newblock \bibinfo{title}{Mapping urban cold-air paths in a central european city using numerical modelling and geospatial analysis}.
\newblock \bibinfo{journal}{Urban Climate} \bibinfo{volume}{29}, \bibinfo{pages}{100503}.
\newblock \URLprefix \url{https://linkinghub.elsevier.com/retrieve/pii/S2212095519300665}, \DOIprefix\doi{10.1016/j.uclim.2019.100503}.
\bibitem[{Heusinger et~al.(2023)Heusinger, Bruchmann and Weber}]{heusinger_modeling_2023}
\bibinfo{author}{Heusinger, J.}, \bibinfo{author}{Bruchmann, N.}, \bibinfo{author}{Weber, S.}, \bibinfo{year}{2023}.
\newblock \bibinfo{title}{Modeling the impacts of building energy efficiency on the thermal microclimate in a midsize german city}.
\newblock \bibinfo{journal}{Urban Climate} \bibinfo{volume}{52}, \bibinfo{pages}{101678}.
\newblock \URLprefix \url{https://linkinghub.elsevier.com/retrieve/pii/S2212095523002729}, \DOIprefix\doi{10.1016/j.uclim.2023.101678}.
\bibitem[{Huang et~al.(2021)Huang, Song, Wang, Chui and Chan}]{huang_synergistic_2021}
\bibinfo{author}{Huang, X.}, \bibinfo{author}{Song, J.}, \bibinfo{author}{Wang, C.}, \bibinfo{author}{Chui, T.F.M.}, \bibinfo{author}{Chan, P.W.}, \bibinfo{year}{2021}.
\newblock \bibinfo{title}{The synergistic effect of urban heat and moisture islands in a compact high-rise city}.
\newblock \bibinfo{journal}{Building and Environment} \bibinfo{volume}{205}, \bibinfo{pages}{108274}.
\newblock \URLprefix \url{https://linkinghub.elsevier.com/retrieve/pii/S0360132321006740}, \DOIprefix\doi{10.1016/j.buildenv.2021.108274}.
\bibitem[{Jendritzky et~al.(2012)Jendritzky, De~Dear and Havenith}]{jendritzky_utciwhy_2012}
\bibinfo{author}{Jendritzky, G.}, \bibinfo{author}{De~Dear, R.}, \bibinfo{author}{Havenith, G.}, \bibinfo{year}{2012}.
\newblock \bibinfo{title}{{UTCI}—why another thermal index?}
\newblock \bibinfo{journal}{International Journal of Biometeorology} \bibinfo{volume}{56}, \bibinfo{pages}{421--428}.
\newblock \URLprefix \url{http://link.springer.com/10.1007/s00484-011-0513-7}, \DOIprefix\doi{10.1007/s00484-011-0513-7}.
\bibitem[{Johra(2021)}]{johra_2021}
\bibinfo{author}{Johra, H.}, \bibinfo{year}{2021}.
\newblock \bibinfo{title}{Thermal properties of building materials - Review and database}.
\newblock Number \bibinfo{number}{289} in \bibinfo{series}{DCE Technical Reports}, \bibinfo{publisher}{Department of the Built Environment, Aalborg University}.
\newblock \DOIprefix\doi{10.54337/aau456230861}.
\bibitem[{Jones and Schonlau(1998)}]{jones_efficient_1998}
\bibinfo{author}{Jones, D.R.}, \bibinfo{author}{Schonlau, M.}, \bibinfo{year}{1998}.
\newblock \bibinfo{title}{Efficient global optimization of expensive black-box functions}.
\newblock \bibinfo{journal}{Journal of Global Optimization} .
\bibitem[{Knowles(2006)}]{knowlesParEGOHybridAlgorithm2006}
\bibinfo{author}{Knowles, J.}, \bibinfo{year}{2006}.
\newblock \bibinfo{title}{{ParEGO}: a hybrid algorithm with on-line landscape approximation for expensive multiobjective optimization problems}.
\newblock \bibinfo{journal}{{IEEE} Transactions on Evolutionary Computation} \bibinfo{volume}{10}, \bibinfo{pages}{50--66}.
\newblock \URLprefix \url{http://ieeexplore.ieee.org/document/1583627/}, \DOIprefix\doi{10.1109/TEVC.2005.851274}.
\bibitem[{Krayenhoff et~al.(2021)Krayenhoff, Broadbent, Zhao, Georgescu, Middel, Voogt, Martilli, Sailor and Erell}]{krayenhoff_cooling_2021}
\bibinfo{author}{Krayenhoff, E.S.}, \bibinfo{author}{Broadbent, A.M.}, \bibinfo{author}{Zhao, L.}, \bibinfo{author}{Georgescu, M.}, \bibinfo{author}{Middel, A.}, \bibinfo{author}{Voogt, J.A.}, \bibinfo{author}{Martilli, A.}, \bibinfo{author}{Sailor, D.J.}, \bibinfo{author}{Erell, E.}, \bibinfo{year}{2021}.
\newblock \bibinfo{title}{Cooling hot cities: a systematic and critical review of the numerical modelling literature}.
\newblock \bibinfo{journal}{Environmental Research Letters} \bibinfo{volume}{16}, \bibinfo{pages}{053007}.
\newblock \URLprefix \url{https://iopscience.iop.org/article/10.1088/1748-9326/abdcf1}, \DOIprefix\doi{10.1088/1748-9326/abdcf1}.
\bibitem[{Kr{\"u}ger(2021)}]{Krueger2021}
\bibinfo{author}{Kr{\"u}ger, E.L.}, \bibinfo{year}{2021}.
\newblock \bibinfo{title}{Literature Review on UTCI Applications}.
\newblock \bibinfo{publisher}{Springer International Publishing}, \bibinfo{address}{Cham}.
\newblock \DOIprefix\doi{10.1007/978-3-030-76716-7_3}.
\bibitem[{Kumar et~al.(2024)Kumar, Debele, Khalili, Halios, Sahani, Aghamohammadi, Andrade, Athanassiadou, Bhui, Calvillo, Cao, Coulon, Edmondson, Fletcher, Dias De~Freitas, Guo, Hort, Katti, Kjeldsen, Lehmann, Locosselli, Malham, Morawska, Parajuli, Rogers, Yao, Wang, Wenk and Jones}]{kumar_urban_2024}
\bibinfo{author}{Kumar, P.}, \bibinfo{author}{Debele, S.E.}, \bibinfo{author}{Khalili, S.}, \bibinfo{author}{Halios, C.H.}, \bibinfo{author}{Sahani, J.}, \bibinfo{author}{Aghamohammadi, N.}, \bibinfo{author}{Andrade, M.D.F.}, \bibinfo{author}{Athanassiadou, M.}, \bibinfo{author}{Bhui, K.}, \bibinfo{author}{Calvillo, N.}, \bibinfo{author}{Cao, S.J.}, \bibinfo{author}{Coulon, F.}, \bibinfo{author}{Edmondson, J.L.}, \bibinfo{author}{Fletcher, D.}, \bibinfo{author}{Dias De~Freitas, E.}, \bibinfo{author}{Guo, H.}, \bibinfo{author}{Hort, M.C.}, \bibinfo{author}{Katti, M.}, \bibinfo{author}{Kjeldsen, T.R.}, \bibinfo{author}{Lehmann, S.}, \bibinfo{author}{Locosselli, G.M.}, \bibinfo{author}{Malham, S.K.}, \bibinfo{author}{Morawska, L.}, \bibinfo{author}{Parajuli, R.}, \bibinfo{author}{Rogers, C.D.}, \bibinfo{author}{Yao, R.}, \bibinfo{author}{Wang, F.}, \bibinfo{author}{Wenk, J.}, \bibinfo{author}{Jones, L.}, \bibinfo{year}{2024}.
\newblock \bibinfo{title}{Urban heat mitigation by green and blue infrastructure: Drivers, effectiveness, and future needs}.
\newblock \bibinfo{journal}{The Innovation} \bibinfo{volume}{5}, \bibinfo{pages}{100588}.
\newblock \URLprefix \url{https://linkinghub.elsevier.com/retrieve/pii/S2666675824000262}, \DOIprefix\doi{10.1016/j.xinn.2024.100588}.
\bibitem[{Kuttler and Weber(2023)}]{kuttler2023}
\bibinfo{author}{Kuttler, W.}, \bibinfo{author}{Weber, S.}, \bibinfo{year}{2023}.
\newblock \bibinfo{title}{Characteristics and phenomena of the urban climate}.
\newblock \bibinfo{journal}{Meteorologische Zeitschrift} \bibinfo{volume}{32}, \bibinfo{pages}{15--47}.
\newblock \URLprefix \url{http://dx.doi.org/10.1127/metz/2023/1153}, \DOIprefix\doi{10.1127/metz/2023/1153}.
\bibitem[{{LGLN}(2024)}]{lgln2024_lod2}
\bibinfo{author}{{LGLN}}, \bibinfo{year}{2024}.
\newblock \bibinfo{title}{3d-gebäudemodell lod2 niedersachsen (opengeodata)}.
\newblock \URLprefix \url{https://ni-lgln-opengeodata.hub.arcgis.com/apps/lgln-opengeodata::3d-gebaeudemodell-lod2/about}. \bibinfo{note}{lizenz: Creative Commons Namensnennung – 4.0 International (CC BY 4.0), \url{http://creativecommons.org/licenses/by/4.0/}. Quelle: LGLN (2024)}.
\bibitem[{Li et~al.(2026)Li, Li, Chen, Zhang, Pauleit and Rahman}]{liStudyParameterCalibration2026}
\bibinfo{author}{Li, Q.}, \bibinfo{author}{Li, Q.}, \bibinfo{author}{Chen, S.}, \bibinfo{author}{Zhang, X.}, \bibinfo{author}{Pauleit, S.}, \bibinfo{author}{Rahman, M.A.}, \bibinfo{year}{2026}.
\newblock \bibinfo{title}{Study on parameter calibration of urban land surface models based on multi-objective bayesian optimization}.
\newblock \bibinfo{journal}{Building and Environment} \bibinfo{volume}{290}, \bibinfo{pages}{114121}.
\newblock \URLprefix \url{https://linkinghub.elsevier.com/retrieve/pii/S0360132325015872}, \DOIprefix\doi{10.1016/j.buildenv.2025.114121}.
\bibitem[{Lo et~al.(2023)Lo, Mitchell, Buzan, Zscheischler, Schneider, Mistry, Kyselý, Lavigne, Da~Silva, Royé, Urban, Armstrong, {Multi‐Country Multi‐City (MCC) Collaborative Research Network}, Gasparrini and Vicedo‐Cabrera}]{lo_optimal_2023}
\bibinfo{author}{Lo, Y.T.E.}, \bibinfo{author}{Mitchell, D.M.}, \bibinfo{author}{Buzan, J.R.}, \bibinfo{author}{Zscheischler, J.}, \bibinfo{author}{Schneider, R.}, \bibinfo{author}{Mistry, M.N.}, \bibinfo{author}{Kyselý, J.}, \bibinfo{author}{Lavigne, {\'E}.}, \bibinfo{author}{Da~Silva, S.P.}, \bibinfo{author}{Royé, D.}, \bibinfo{author}{Urban, A.}, \bibinfo{author}{Armstrong, B.}, \bibinfo{author}{{Multi‐Country Multi‐City (MCC) Collaborative Research Network}}, \bibinfo{author}{Gasparrini, A.}, \bibinfo{author}{Vicedo‐Cabrera, A.M.}, \bibinfo{year}{2023}.
\newblock \bibinfo{title}{Optimal heat stress metric for modelling heat‐related mortality varies from country to country}.
\newblock \bibinfo{journal}{International Journal of Climatology} \bibinfo{volume}{43}, \bibinfo{pages}{5553--5568}.
\newblock \URLprefix \url{https://rmets.onlinelibrary.wiley.com/doi/10.1002/joc.8160}, \DOIprefix\doi{10.1002/joc.8160}.
\bibitem[{Lüthi et~al.(2023)Lüthi, Fairless, Fischer, Scovronick, {Ben Armstrong}, Coelho, Guo, Guo, Honda, Huber, Kyselý, Lavigne, Royé, Ryti, Silva, Urban, Gasparrini, Bresch and Vicedo-Cabrera}]{luthi_rapid_2023}
\bibinfo{author}{Lüthi, S.}, \bibinfo{author}{Fairless, C.}, \bibinfo{author}{Fischer, E.M.}, \bibinfo{author}{Scovronick, N.}, \bibinfo{author}{{Ben Armstrong}}, \bibinfo{author}{Coelho, M.D.S.Z.S.}, \bibinfo{author}{Guo, Y.L.}, \bibinfo{author}{Guo, Y.}, \bibinfo{author}{Honda, Y.}, \bibinfo{author}{Huber, V.}, \bibinfo{author}{Kyselý, J.}, \bibinfo{author}{Lavigne, {\'E}.}, \bibinfo{author}{Royé, D.}, \bibinfo{author}{Ryti, N.}, \bibinfo{author}{Silva, S.}, \bibinfo{author}{Urban, A.}, \bibinfo{author}{Gasparrini, A.}, \bibinfo{author}{Bresch, D.N.}, \bibinfo{author}{Vicedo-Cabrera, A.M.}, \bibinfo{year}{2023}.
\newblock \bibinfo{title}{Rapid increase in the risk of heat-related mortality}.
\newblock \bibinfo{journal}{Nature Communications} \bibinfo{volume}{14}, \bibinfo{pages}{4894}.
\newblock \URLprefix \url{https://www.nature.com/articles/s41467-023-40599-x}, \DOIprefix\doi{10.1038/s41467-023-40599-x}.
\bibitem[{Macasieb et~al.(2024)Macasieb, White, Pasetto and Siade}]{macasieb_probabilistic_2024}
\bibinfo{author}{Macasieb, R.Q.}, \bibinfo{author}{White, J.}, \bibinfo{author}{Pasetto, D.}, \bibinfo{author}{Siade, A.J.}, \bibinfo{year}{2024}.
\newblock \bibinfo{title}{A probabilistic approach to surrogate assisted multi-objective optimization of complex groundwater problems}.
\newblock \URLprefix \url{https://essopenarchive.org/doi/full/10.22541/essoar.172408236.65180534/v1}, \DOIprefix\doi{10.22541/essoar.172408236.65180534/v1}.
\bibitem[{Markolf and Weber(2026)}]{markolf-2026_NEE_greenRoof}
\bibinfo{author}{Markolf, N.}, \bibinfo{author}{Weber, S.}, \bibinfo{year}{2026}.
\newblock \bibinfo{title}{Net ecosystem exchange of extensive green roofs: the role of coupled energy, carbon, and water fluxes quantified by long-term micrometeorological observations}.
\newblock \bibinfo{journal}{Biogeosciences} \bibinfo{volume}{23}, \bibinfo{pages}{5035--5053}.
\newblock \URLprefix \url{https://bg.copernicus.org/articles/23/5035/2026/}, \DOIprefix\doi{10.5194/bg-23-5035-2026}.
\bibitem[{Masselot et~al.(2025)Masselot, Mistry, Rao, Huber, Monteiro, Samoli, Stafoggia, de’Donato, Garcia-Leon, Ciscar, Feyen, Schneider, Katsouyanni, Vicedo-Cabrera, Aunan and Gasparrini}]{masselot_estimating_2025}
\bibinfo{author}{Masselot, P.}, \bibinfo{author}{Mistry, M.N.}, \bibinfo{author}{Rao, S.}, \bibinfo{author}{Huber, V.}, \bibinfo{author}{Monteiro, A.}, \bibinfo{author}{Samoli, E.}, \bibinfo{author}{Stafoggia, M.}, \bibinfo{author}{de’Donato, F.}, \bibinfo{author}{Garcia-Leon, D.}, \bibinfo{author}{Ciscar, J.C.}, \bibinfo{author}{Feyen, L.}, \bibinfo{author}{Schneider, A.}, \bibinfo{author}{Katsouyanni, K.}, \bibinfo{author}{Vicedo-Cabrera, A.M.}, \bibinfo{author}{Aunan, K.}, \bibinfo{author}{Gasparrini, A.}, \bibinfo{year}{2025}.
\newblock \bibinfo{title}{Estimating future heat-related and cold-related mortality under climate change, demographic and adaptation scenarios in 854 european cities}.
\newblock \bibinfo{journal}{Nature Medicine} \bibinfo{volume}{31}, \bibinfo{pages}{1294--1302}.
\newblock \URLprefix \url{https://www.nature.com/articles/s41591-024-03452-2}, \DOIprefix\doi{10.1038/s41591-024-03452-2}.
\bibitem[{Masson(2000)}]{masson_physically-based_2000}
\bibinfo{author}{Masson, V.}, \bibinfo{year}{2000}.
\newblock \bibinfo{title}{A {Physically}-{Based} {Scheme} {For} {The} {Urban} {Energy} {Budget} {In} {Atmospheric} {Models}}.
\newblock \bibinfo{journal}{Boundary-Layer Meteorology} \bibinfo{volume}{94}, \bibinfo{pages}{357--397}.
\newblock \URLprefix \url{https://link.springer.com/10.1023/A:1002463829265}, \DOIprefix\doi{10.1023/A:1002463829265}.
\bibitem[{Mateen et~al.(2025)Mateen, Giometto, Biswal, Tapper and Parlange}]{mateen_large_2025}
\bibinfo{author}{Mateen, S.A.}, \bibinfo{author}{Giometto, M.G.}, \bibinfo{author}{Biswal, B.}, \bibinfo{author}{Tapper, N.J.}, \bibinfo{author}{Parlange, M.B.}, \bibinfo{year}{2025}.
\newblock \bibinfo{title}{Large eddy simulation based evaluation of an urban canopy model}.
\newblock \bibinfo{journal}{Boundary-Layer Meteorology} \bibinfo{volume}{191}, \bibinfo{pages}{34}.
\newblock \URLprefix \url{https://link.springer.com/10.1007/s10546-025-00927-8}, \DOIprefix\doi{10.1007/s10546-025-00927-8}.
\bibitem[{Meili et~al.(2020a)Meili, Manoli, Burlando, Bou-Zeid, Chow, Coutts, Daly, Nice, Roth, Tapper, Velasco, Vivoni and Fatichi}]{meili_supplement_2020}
\bibinfo{author}{Meili, N.}, \bibinfo{author}{Manoli, G.}, \bibinfo{author}{Burlando, P.}, \bibinfo{author}{Bou-Zeid, E.}, \bibinfo{author}{Chow, W.T.L.}, \bibinfo{author}{Coutts, A.M.}, \bibinfo{author}{Daly, E.}, \bibinfo{author}{Nice, K.A.}, \bibinfo{author}{Roth, M.}, \bibinfo{author}{Tapper, N.J.}, \bibinfo{author}{Velasco, E.}, \bibinfo{author}{Vivoni, E.R.}, \bibinfo{author}{Fatichi, S.}, \bibinfo{year}{2020}a.
\newblock \bibinfo{title}{Supplement of an urban ecohydrological model to quantify the effect of vegetation on urban climate and hydrology ({UT}\&c v1.0)}.
\newblock \DOIprefix\doi{https://doi.org/10.5194/gmd-13-335-2020-supplement}.
\bibitem[{Meili et~al.(2020b)Meili, Manoli, Burlando, Bou-Zeid, Chow, Coutts, Daly, Nice, Roth, Tapper, Velasco, Vivoni and Fatichi}]{meili2020utc}
\bibinfo{author}{Meili, N.}, \bibinfo{author}{Manoli, G.}, \bibinfo{author}{Burlando, P.}, \bibinfo{author}{Bou-Zeid, E.}, \bibinfo{author}{Chow, W.T.L.}, \bibinfo{author}{Coutts, A.M.}, \bibinfo{author}{Daly, E.}, \bibinfo{author}{Nice, K.A.}, \bibinfo{author}{Roth, M.}, \bibinfo{author}{Tapper, N.J.}, \bibinfo{author}{Velasco, E.}, \bibinfo{author}{Vivoni, E.R.}, \bibinfo{author}{Fatichi, S.}, \bibinfo{year}{2020}b.
\newblock \bibinfo{title}{An urban ecohydrological model to quantify the effect of vegetation on urban climate and hydrology (ut\&c v1.0)}.
\newblock \bibinfo{journal}{Geoscientific Model Development} \bibinfo{volume}{13}, \bibinfo{pages}{335--362}.
\newblock \URLprefix \url{https://gmd.copernicus.org/articles/13/335/2020/}, \DOIprefix\doi{10.5194/gmd-13-335-2020}.
\bibitem[{Meili et~al.(2021)Meili, Manoli, Burlando, Carmeliet, Chow, Coutts, Velasco, Vivoni and Fatichi}]{meili2021trees}
\bibinfo{author}{Meili, N.}, \bibinfo{author}{Manoli, G.}, \bibinfo{author}{Burlando, P.}, \bibinfo{author}{Carmeliet, J.}, \bibinfo{author}{Chow, W.T.L.}, \bibinfo{author}{Coutts, Andrew M.and~Roth, M.}, \bibinfo{author}{Velasco, E.}, \bibinfo{author}{Vivoni, E.R.}, \bibinfo{author}{Fatichi, S.}, \bibinfo{year}{2021}.
\newblock \bibinfo{title}{Tree effects on urban microclimate: Diurnal, seasonal, and climatic temperature differences explained by separating radiation, evapotranspiration, and roughness effects}.
\newblock \bibinfo{journal}{Urban Forestry \& Urban Greening} \bibinfo{volume}{58}, \bibinfo{pages}{126970}.
\newblock \URLprefix \url{https://www.sciencedirect.com/science/article/pii/S1618866720307871}, \DOIprefix\doi{10.1016/j.ufug.2020.126970}.
\bibitem[{Meili et~al.(2025)Meili, Zheng, Takane, Nakajima, Yamaguchi, Chi, Zhu, Wang, Qiu, Paschalis, Manoli, Burlando, Tan and Fatichi}]{meili_modeling_2025}
\bibinfo{author}{Meili, N.}, \bibinfo{author}{Zheng, X.}, \bibinfo{author}{Takane, Y.}, \bibinfo{author}{Nakajima, K.}, \bibinfo{author}{Yamaguchi, K.}, \bibinfo{author}{Chi, D.}, \bibinfo{author}{Zhu, Y.}, \bibinfo{author}{Wang, J.}, \bibinfo{author}{Qiu, Y.}, \bibinfo{author}{Paschalis, A.}, \bibinfo{author}{Manoli, G.}, \bibinfo{author}{Burlando, P.}, \bibinfo{author}{Tan, P.Y.}, \bibinfo{author}{Fatichi, S.}, \bibinfo{year}{2025}.
\newblock \bibinfo{title}{Modeling the effect of trees on energy demand for indoor cooling and dehumidification across cities and climates}.
\newblock \bibinfo{journal}{Journal of Advances in Modeling Earth Systems} \bibinfo{volume}{17}, \bibinfo{pages}{e2024MS004590}.
\newblock \URLprefix \url{https://agupubs.onlinelibrary.wiley.com/doi/10.1029/2024MS004590}, \DOIprefix\doi{10.1029/2024MS004590}.
\bibitem[{Münzinger(2025)}]{muenzinger2025semantic}
\bibinfo{author}{Münzinger, M.}, \bibinfo{year}{2025}.
\newblock \bibinfo{title}{Lidar-based tree models for braunschweig, germany (2019)}.
\newblock \URLprefix \url{https://doi.org/10.71830/2DZCIA}, \DOIprefix\doi{10.71830/2DZCIA}.
\bibitem[{Oke(1973)}]{OKE1973769}
\bibinfo{author}{Oke, T.}, \bibinfo{year}{1973}.
\newblock \bibinfo{title}{City size and the urban heat island}.
\newblock \bibinfo{journal}{Atmospheric Environment (1967)} \bibinfo{volume}{7}, \bibinfo{pages}{769--779}.
\newblock \URLprefix \url{https://www.sciencedirect.com/science/article/pii/0004698173901406}, \DOIprefix\doi{https://doi.org/10.1016/0004-6981(73)90140-6}.
\bibitem[{Oke et~al.(2017)Oke, Mills, Christen and Voogt}]{Oke_Mills_Christen_Voogt_2017}
\bibinfo{author}{Oke, T.R.}, \bibinfo{author}{Mills, G.}, \bibinfo{author}{Christen, A.}, \bibinfo{author}{Voogt, J.A.}, \bibinfo{year}{2017}.
\newblock \bibinfo{title}{Urban Climates}.
\newblock \bibinfo{publisher}{Cambridge University Press}.
\bibitem[{Rasmussen and Williams(2006)}]{rasmussenGaussianProcessesMachine2008}
\bibinfo{author}{Rasmussen, C.E.}, \bibinfo{author}{Williams, C.K.I.}, \bibinfo{year}{2006}.
\newblock \bibinfo{title}{Gaussian processes for machine learning}.
\newblock Adaptive computation and machine learning. \bibinfo{edition}{3. print} ed., \bibinfo{publisher}{{MIT} Press}.
\bibitem[{{Robert Koch-Institut}(2026)}]{robert_koch-institut_wochenbericht_2026}
\bibinfo{author}{{Robert Koch-Institut}}, \bibinfo{year}{2026}.
\newblock \bibinfo{title}{Wochenbericht zur hitzebedingten mortalität {KW} 26} \URLprefix \url{https://edoc.rki.de/handle/176904/13783}, \DOIprefix\doi{10.25646/14292}. \bibinfo{note}{publisher: Robert Koch-Institut}.
\bibitem[{Saltelli et~al.(2010)Saltelli, Annoni, Azzini, Campolongo, Ratto and Tarantola}]{saltelli_variance_2010}
\bibinfo{author}{Saltelli, A.}, \bibinfo{author}{Annoni, P.}, \bibinfo{author}{Azzini, I.}, \bibinfo{author}{Campolongo, F.}, \bibinfo{author}{Ratto, M.}, \bibinfo{author}{Tarantola, S.}, \bibinfo{year}{2010}.
\newblock \bibinfo{title}{Variance based sensitivity analysis of model output. design and estimator for the total sensitivity index}.
\newblock \bibinfo{journal}{Computer Physics Communications} \bibinfo{volume}{181}, \bibinfo{pages}{259--270}.
\newblock \URLprefix \url{https://linkinghub.elsevier.com/retrieve/pii/S0010465509003087}, \DOIprefix\doi{10.1016/j.cpc.2009.09.018}.
\bibitem[{Schrijvers et~al.(2016)Schrijvers, Jonker, De~Roode and Kenjereš}]{schrijvers_effect_2016}
\bibinfo{author}{Schrijvers, P.}, \bibinfo{author}{Jonker, H.}, \bibinfo{author}{De~Roode, S.}, \bibinfo{author}{Kenjereš, S.}, \bibinfo{year}{2016}.
\newblock \bibinfo{title}{The effect of using a high-albedo material on the universal temperature climate index within a street canyon}.
\newblock \bibinfo{journal}{Urban Climate} \bibinfo{volume}{17}, \bibinfo{pages}{284--303}.
\newblock \URLprefix \url{https://linkinghub.elsevier.com/retrieve/pii/S2212095516300104}, \DOIprefix\doi{10.1016/j.uclim.2016.02.005}.
\bibitem[{Shen et~al.(2026)Shen, Gao, Liu, Zhang and Zheng}]{shen_parametric_2026}
\bibinfo{author}{Shen, P.}, \bibinfo{author}{Gao, X.}, \bibinfo{author}{Liu, Y.}, \bibinfo{author}{Zhang, Y.}, \bibinfo{author}{Zheng, X.}, \bibinfo{year}{2026}.
\newblock \bibinfo{title}{Parametric optimization of urban residential morphology for outdoor thermal comfort with integrated code regulatory and site development requirement}.
\newblock \bibinfo{journal}{Urban Climate} \bibinfo{volume}{65}, \bibinfo{pages}{102836}.
\newblock \URLprefix \url{https://linkinghub.elsevier.com/retrieve/pii/S2212095526000672}, \DOIprefix\doi{10.1016/j.uclim.2026.102836}.
\bibitem[{{Stadt Braunschweig}(2026)}]{braunschweig_prinzpark}
\bibinfo{author}{{Stadt Braunschweig}}, \bibinfo{year}{2026}.
\newblock \bibinfo{title}{Prinz-albrecht-park}.
\newblock \URLprefix \url{https://www.braunschweig.de/leben/im_gruenen/gruenanlagen/PrinzPark.php}. \bibinfo{note}{accessed: 2026-05-29}.
\bibitem[{Stewart and Oke(2012)}]{stewart_local_2012}
\bibinfo{author}{Stewart, I.D.}, \bibinfo{author}{Oke, T.R.}, \bibinfo{year}{2012}.
\newblock \bibinfo{title}{Local climate zones for urban temperature studies}.
\newblock \bibinfo{journal}{Bulletin of the American Meteorological Society} \bibinfo{volume}{93}, \bibinfo{pages}{1879--1900}.
\newblock \URLprefix \url{https://journals.ametsoc.org/doi/10.1175/BAMS-D-11-00019.1}, \DOIprefix\doi{10.1175/BAMS-D-11-00019.1}.
\bibitem[{Tan et~al.(2020)Tan, Wong, Tan, Jusuf, Schmiele and Chiam}]{tan_transpiration_2020}
\bibinfo{author}{Tan, P.Y.}, \bibinfo{author}{Wong, N.H.}, \bibinfo{author}{Tan, C.L.}, \bibinfo{author}{Jusuf, S.K.}, \bibinfo{author}{Schmiele, K.}, \bibinfo{author}{Chiam, Z.Q.}, \bibinfo{year}{2020}.
\newblock \bibinfo{title}{Transpiration and cooling potential of tropical urban trees from different native habitats}.
\newblock \bibinfo{journal}{Science of The Total Environment} \bibinfo{volume}{705}, \bibinfo{pages}{135764}.
\newblock \URLprefix \url{https://linkinghub.elsevier.com/retrieve/pii/S0048969719357596}, \DOIprefix\doi{10.1016/j.scitotenv.2019.135764}.
\bibitem[{United~Nations(2025)}]{UN2025Urbanization}
\bibinfo{author}{United~Nations, P.D.}, \bibinfo{year}{2025}.
\newblock \bibinfo{title}{World Urbanization Prospects 2025: Summary of Results}.
\newblock \bibinfo{type}{UN DESA/POP/2025/TR/NO. 12}. United Nations, Department of Economic and Social Affairs, Population Division. \bibinfo{address}{New York, NY, USA}.
\newblock \URLprefix \url{https://population.un.org/wup/assets/Publications/undesa\_pd\_2025\_wup2025\_summary\_of\_results\_final.pdf}.
\bibitem[{Walter(2026a)}]{MyCode2026}
\bibinfo{author}{Walter, R.}, \bibinfo{year}{2026}a.
\newblock \bibinfo{title}{MOBO framework for urban heat mitigation using UT\&C}.
\newblock \DOIprefix\doi{10.5281/zenodo.22706138}. \bibinfo{note}{code will be published upon acceptance}.
\bibitem[{Walter(2026b)}]{walter_respace_interaction_2026}
\bibinfo{author}{Walter, R.}, \bibinfo{year}{2026}b.
\newblock \bibinfo{title}{RespaceInteraction: Urban climate design explorer}.
\newblock \URLprefix \url{https://github.com/Rebwalter/ReSpaceInteraction}. \bibinfo{note}{software repository}.
\bibitem[{Wujeska-Klause and Pfautsch(2020)}]{wujeska_2020_nighttimecooling}
\bibinfo{author}{Wujeska-Klause, A.}, \bibinfo{author}{Pfautsch, S.}, \bibinfo{year}{2020}.
\newblock \bibinfo{title}{The best urban trees for daytime cooling leave nights slightly warmer}.
\newblock \bibinfo{journal}{Forests} \bibinfo{volume}{11}.
\newblock \URLprefix \url{https://www.mdpi.com/1999-4907/11/9/945}, \DOIprefix\doi{10.3390/f11090945}.
\bibitem[{Zhao et~al.(2023)Zhao, Li, Bardhan, Kubilay, Li and Carmeliet}]{zhao_time-evolving_tree_2023}
\bibinfo{author}{Zhao, Y.}, \bibinfo{author}{Li, H.}, \bibinfo{author}{Bardhan, R.}, \bibinfo{author}{Kubilay, A.}, \bibinfo{author}{Li, Q.}, \bibinfo{author}{Carmeliet, J.}, \bibinfo{year}{2023}.
\newblock \bibinfo{title}{The time-evolving impact of tree size on nighttime street canyon microclimate: Wind tunnel modeling of aerodynamic effects and heat removal}.
\newblock \bibinfo{journal}{Urban Climate} \bibinfo{volume}{49}, \bibinfo{pages}{101528}.
\newblock \URLprefix \url{https://linkinghub.elsevier.com/retrieve/pii/S2212095523001220}, \DOIprefix\doi{10.1016/j.uclim.2023.101528}.

\end{thebibliography}
\end{document}